\documentclass[aps,pre,twocolumn,amsmath,amssymb,superscriptaddress,floatfix,citeautoscript]{revtex4-1}

\usepackage{graphicx, multirow, xcolor, bm, ulem}
\usepackage{amsfonts,amssymb,amsmath}
\usepackage{graphicx,dcolumn,bm,color,braket,slashed}
\usepackage{times, comment, mathtools, extarrows}
\usepackage{hyperref}
\hypersetup{colorlinks=true, citecolor=blue, urlcolor=blue, linkcolor=blue}

\def\ket#1{|#1\rangle }
\def\Ket#1{\biggl|#1\biggr\rangle }
\def\bra#1{\langle #1 |}

\newcommand\abs[1]{\left|#1\right|}

\begin{document}

\title{
Topological features of ground states and topological solitons in generalized Su-Schrieffer-Heeger models using generalized time-reversal, particle-hole, and chiral symmetries
}

\author{Sang-Hoon Han}
\thanks{These authors contributed equally to this work.}
\affiliation{Department of Physics, Hanyang University, Seoul 04763, Korea}

\author{Seung-Gyo Jeong}
\thanks{These authors contributed equally to this work.}
\affiliation{Department of Physics, Pohang University of Science and Technology (POSTECH), Pohang, 37673, Korea}
\affiliation{Center for Artificial Low Dimensional Electronic Systems, Institute for Basic Science (IBS), Pohang, 37673, Korea}

\author{Sun-Woo Kim}
\affiliation{Department of Physics, Hanyang University, Seoul 04763, Korea}

\author{Tae-Hwan Kim}
\email{taehwan@postech.ac.kr}
\affiliation{Department of Physics, Pohang University of Science and Technology (POSTECH), Pohang, 37673, Korea}

\author{Sangmo Cheon}
\email{sangmocheon@hanyang.ac.kr}
\affiliation{Department of Physics, Hanyang University, Seoul 04763, Korea}
\affiliation{Research Institute for Natural Science, Hanyang University, Seoul 04763, Korea}

\begin{abstract}
Topological phases and their topological features are enriched by the fundamental time-reversal, particle-hole, and chiral as well as crystalline symmetries.
While one-dimensional (1D) generalized Su-Schrieffer-Heeger (SSH) systems show various topological phenomena such as topological solitons and topological charge pumping, 
it remains unclear how such symmetry protects and relates such topological phenomena.
Here we show that the generalized time-reversal, particle-hole, and chiral symmetry operators consistently explain  not  only  the  symmetry  transformation  properties  between the  ground states  but  also  the  topological  features  of the topological solitons in prototypical quasi-1D systems such as the SSH, Rice-Mele, and double-chain models.
As a consequence, we classify generalized essential operators into three groups:
Class~I and class II operators connect ground states in between after spontaneous symmetry breaking while class~III operators give the generalized particle-hole and chiral symmetries to ground states.
Furthermore, class~I operators endow the equivalence relation between  topological solitons while class~II and III operators do the particle-hole relations.
Finally, we demonstrate three distinct types of topological charge pumping and soliton chirality from the viewpoint of class~I, II, and III operators. 
We build a general framework to explore the topological features of the generalized 1D electronic system, which can be easily applied in various condensed  matter  systems  as  well  as  photonic  crystal  and  cold atomic systems.
\end{abstract}

\maketitle

\section{Introduction}
CPT symmetries are fundamental symmetries in nature and play important roles in the CPT theorem.
In condensed matter systems, 
time-reversal $\hat{\mathcal{T}}$, 
charge-conjugation $\hat{\mathcal{C}}$ (or particle-hole),
and chiral  $\hat{\Gamma} \equiv  \hat{\mathcal{T}}\hat{\mathcal{C}} $
symmetry operators are
the fundamental symmetry operators for not only classifying the topological insulators and superconductors
but also endowing various symmetries and dualities to the quasiparticles
within condensed matter CPT theorem~\cite{schnyder2008, hsieh2014, chiu2016}.
For example, quantum spin Hall insulator~\cite{kane2005, hasan2010, bernevig2006} and Majorana fermion~\cite{kitaev2001, elliott2015} are protected by time-reversal and particle-hole symmetry, respectively,
and a skyrmion-antiskyrmion pair satisfies the particle-hole relations via $\hat{\mathcal{T}}$ allowing pair creation and pair annihilation~\cite{romming2013, koshibae2016, stier2017}.
%

As one of the most famous one-dimensional (1D) topological insulators, 
the Su-Schrieffer-Heeger~\cite{SSH1979, SSH1988} model exhibits fascinating topological phenomena 
such as topologically nontrivial ground states, topological Jackiw-Rebbi solitons, fractional fermion number, and spin-charge separation~\cite{jackiw1976, goldstone1981, jackiw1981}.
Such topological features are protected by time-reversal, particle-hole, and chiral symmetries leading to the BDI class~\cite{schnyder2008}.
Beyond the Su-Schrieffer-Heeger (SSH) system,
the Rice-Mele (RM)~\cite{RM1982}, the extended SSH~\cite{Fu2006, Wang2013}, and double-chain (DC)~\cite{cheon2015} systems introduce more interesting topological solitons and topological Thouless charge pumping by manipulating symmetries~\cite{thouless1983, lohse2016, goldman2016, lu2016, nakajima2016}.
In particular, 
chirality of topological solitons in the DC system emerges and such solitons are named as chiral solitons~\cite{cheon2015}.

However, such topological features of the extended quasi-1D electronic systems have not been studied yet in terms of $\hat{\mathcal{T}}$, $\hat{\mathcal{C}}$, and $\hat{\Gamma}$.
In fact, only $\hat{\mathcal{T}}$, $\hat{\mathcal{C}}$, and $\hat{\Gamma}$  cannot support topological features of extended 1D electronic systems
without considering the discreteness of lattice systems.
For example, nonsymmorphic symmetries may enforce the existence of topological band crossings in the bulk of 1D system~\cite{Zhao2016}.
Therefore, additional proper operators reflecting the entire symmetry of discrete lattice systems are required to understand various topological features.
One of the purposes of this work is to give a general framework that consistently explains the topological features of extended 1D electronic systems
by using $\hat{\mathcal{T}}$, $\hat{\mathcal{C}}$, $\hat{\Gamma}$, nonsymmorphic, and their composite operators.

On the other hand, 
spontaneous symmetry breaking and the Nambu-Goldstone theorem~\cite{nambu1960,goldstone1962} are fundamental principles
that stipulate the fundamental phenomena of diverse symmetry-broken systems, 
from the ferromagnetism and superconductivity in condensed matter physics 
to the Higgs mechanism in particle physics.
In a 1D electronic topological system,
after the spontaneous symmetry breaking of Peierls dimerization~\cite{gruner1988} occurs,
a remaining crystalline symmetry (such as inversion) gives the topological classification of energetically degenerate but distinct ground states~\cite{kane2013, Hughes2011}.
At the same time, the broken symmetry operators relate the degenerate ground states,
which is similar to the Nambu-Goldstone theorem for the continuous system.
Due to the lack of continuous symmetry in the lattice systems,
instead of massless Goldstone bosons, 
topological solitons with finite excitation energy emerge
and the broken symmetry operators may endow unique relations to the ground states as well as topological solitons.
However, no systematic study reveals the interplay between spontaneous  symmetry  breaking and topology.
Here we generalize the Goldstone theorem
to explain not only the symmetry transformation properties between the ground states but also 
the various dualities between the topological solitons in prototypical quasi-1D systems.

In this work, we show that the generalized time-reversal, particle-hole, and chiral symmetry operators consistently explain not only the symmetry transformation properties between the ground states but also the topological features of the topological solitons in prototypical quasi-1D systems such as the SSH, RM, and DC models.
Combining fundamental symmetry operators ($\hat{\mathcal{T}}, \hat{\mathcal{C}}$, and $\hat \Gamma$) and nonsymmorphic crystalline operators,
we establish three classes of essential symmetry operators according to their natures and roles (see Table~\ref{appendix:table:CTP_extended}).
The class~I and II operators connect distinct ground states after spontaneous symmetry breaking 
while the class~III operators give particle-hole and chiral symmetries regardless of spontaneous symmetry breaking.
The class~I (II and III) operators endow the equivalence (particle-hole) relation between ground states as well as topological solitons.
Using class~I, II, and III operators, we derive the topological properties of topological solitons and their $\mathbb{Z}_2$ or $\mathbb{Z}_4$ group structures.
Furthermore, we systematically demonstrate three distinct types of topological charge pumping and soliton chirality in the SSH, RM, and DC models.
We build a general framework to explore the topological features of the generalized 1D electronic system, which can be easily applied in various condensed matter systems as well as photonic crystal~\cite{ozawa2019} and cold atomic systems~\cite{cooper2019}.

\begin{table}[tb]
\caption{
\label{appendix:table:CTP_extended}
Symmetry properties of the SSH, RM, and DC models under the class~I ($\hat{\mathcal{O}}_{\text{I}}$), II ($\hat{\mathcal{C}}_{\text{II}}$, $\hat{\Gamma}_{\text{II}}$), and III ($\hat{\mathcal{C}}_{\text{III}}$, $\hat{\Gamma}_{\text{III}}$) operators for both undimerized and dimerized phases.
The presence and absence of symmetries is denoted by $\pm 1$ and $0$, respectively.
$+1$ or $-1$ denotes the square of the corresponding operator.
When the symmetry properties are distinct for undimerized and dimerized phases, their properties are indicated without (with) parentheses for dimerized (undimerized) phases.
}
{\renewcommand{\arraystretch}{0.6}%
\begin{tabular}{@{\extracolsep{8pt}} l  cccccc  }
\hline
\hline
\\
& &
\multicolumn{1}{c}{Class~I}~& \multicolumn{2}{c}{Class~II} & \multicolumn{2}{c}{~~Class~III}
\\
\\
\cline{3-3} \cline{4-5} \cline{6-7}
\\
Model & ~~~$\hat{\mathcal{T}}$~~~ & ~~~$\hat{\mathcal{O}}_{\text{I}}$~~~ & ~~~$\hat{\mathcal{C}}_{\text{II}}$~~~ & ~$\hat{\Gamma}_{\text{II}}$~~ & ~~~~~$\hat{\mathcal{C}}_{\text{III}}$~~ & ~~~~$\hat{\Gamma}_{\text{III}}$~ \\
\\
\hline 
\\
SSH & 1 & 0 (1) & 0 (1) & 0 (1) & \phantom{$-$}1 & \phantom{$-$}1\\ 
RM & 1 & 0 (1) & 0 (1) & 0 (1) & \phantom{$-$}1 & \phantom{$-$}1\\
DC & 1 & 0 (1) & 0 (1) & 0 (1) & $-1$ & $-1$ \\
\\
\hline
\hline
\end{tabular}
}
\end{table}

\begin{figure}[t]
\centering \includegraphics[width=86mm]{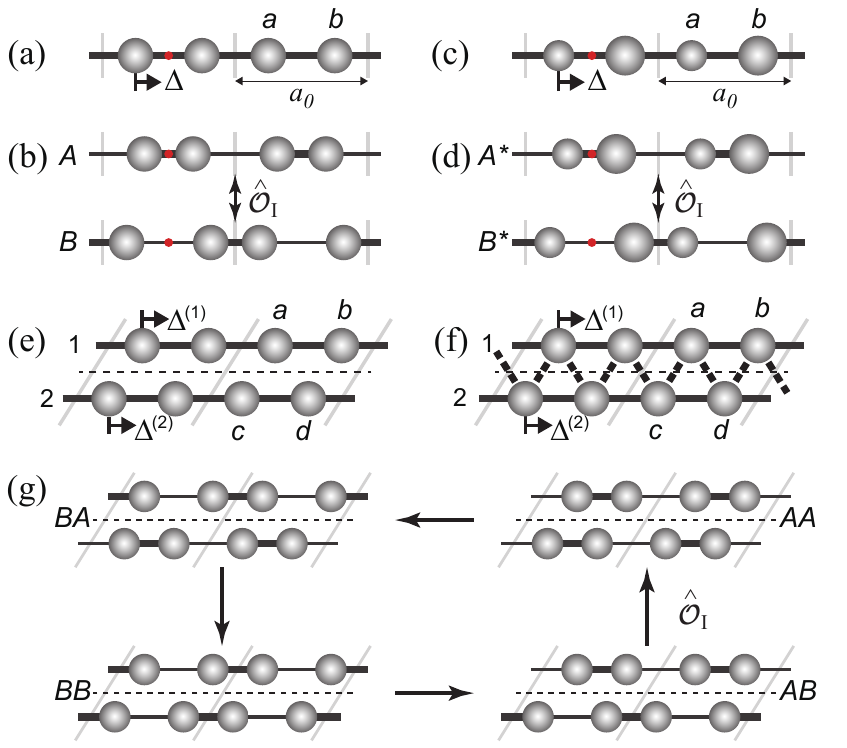} 
\caption{ \label{fig:models}
[(a) and (b)] SSH model.
[(c) and (d)] RM model.
Panels (a) and (c) [panels (b) and (d)] represent undimerized (dimerized) phases.
$\Delta$ is the dimerization displacement and $a$ and $b$ label the sublattice atoms in the unit cell.
Each unit cell is indicated by gray solid lines.
In (b) [(d)], $A$ ($A^*$) and $B$ ($B^*$) are two degenerate dimerized phases in the SSH (RM) model, respectively,
and class~I operator $\hat{\mathcal{O}_\text{I}}$ connects them. 
In (b), $\hat{\mathcal{O}_\text{I}} = \hat{G} \equiv  \{ E | \frac{a_0}{2} \}$
is the half-translation operator while, in (d),
$\hat{\mathcal{O}_\text{I}} = \hat{G}^{\text{PT}} \equiv \hat{G} \hat{\mathcal{P}}\hat{\mathcal{T}}$ is the parity-time-symmetric half-translation operator.
Red circles indicate the inversion centers for the inversion operator $\hat{\mathcal{P}}$.
[(e)--(g)] DC model.
Undimerized double chains (e) without and (f) with interchain coupling which is indicated by zigzag dashed lines.
$\Delta^{(i=1,2)}$ indicates the dimerization displacement of the $i$th chain 
and $a$, $b$, $c$, and $d$ denote the sublattice atoms.
(g) Four dimerized phases, where the zigzag dashed lines are omitted for simplicity.
The class~I operator ($\hat{\mathcal{O}_\text{I}}$) cyclically connects the four dimerized phases.
In (g), $\hat{\mathcal{O}_\text{I}}  = \hat{G}_y \equiv  \{ M_y | -\frac{a_0}{4} \}$
is the glide reflection symmetry operator,
where $M_y$ is a reflection operator with respect to the $xz$ plane (horizontal dashed lines).
} 
\end{figure}

\section{Models}
In this work, we consider three concrete models: the SSH~\cite{SSH1979}, RM~\cite{RM1982}, and DC models~\cite{cheon2015}.
For each model, we construct a tight-binding Hamiltonian $H$, Bloch Hamiltonian $\mathcal{H}$, and low-energy effective Hamiltonian $\mathsf{H}$ to investigate the symmetry relations and topological features of the ground states as well as topological solitons.

\subsection{Single-chain model}
The SSH [Fig.~\ref{fig:models}(a)] and RM models [Fig.~\ref{fig:models}(c)] are basic building blocks in 1D electronic systems.
The SSH (RM) model has two identical (different) atoms without (with) a staggered sublattice potential $m_z $ in the unit cell.
For both model, we construct a general single-chain tight-binding Hamiltonian $H_{\text{single}}$
which is composed of 
the electron hopping Hamiltonian $H_{ e  }$ between two atoms,
the onsite Hamiltonian $H_{ \text{on}  }$ for the staggered sublattice potential,
and the phonon Hamiltonian $H_{ \text{ph} }$:
\begin{eqnarray}
	&& H_{ \text{single} } =  H_{e} + H_{\text{on}} + H_{\text{ph}} , \label{eq:Single_1}  \\
	&& H_{ e  }  =  \sum_{n,s} t_{n+1,n} c^{\dagger}_{n+1,s} c_{n, s}  + \text{H.c.} , \nonumber \\
	&& H_{\text{on}} =  \sum_{n=1}^{N/2} m_z \left( c^{\dagger}_{2n-1, s} c_{2n-1, s} - c^{\dagger}_{2n, s} c_{2n, s} \right ),  \nonumber \\
	&& H_{ \text{ph} } =  \sum_{n=1}^{N} \frac{1}{2} M_n  {\dot u_n}^2 + \frac{1}{2} K (u_{n+1} - u_{n} )^2,  \nonumber
\end{eqnarray}
where $t_{n+1,n}$ is the nearest hopping integral from the $n$th site to the $(n+1)$th site,
$c^{\dagger}_{n, s}$ ($c_{n, s}$)  is the creation (annihilation) operator of an electron with spin $s$ on the $n$th site.
$N$ is the total number of atoms,
$K$ is the harmonic spring constant when expanded to the second order about the undimerized phase,
and $u_n$ is the displacement of the $n$th atom.
$M_{2n+1} = M_a$ and $M_{2n} = M_b$ and $M_a$ and $M_b$ are the masses of two atoms in a unit cell.
The electron-phonon interaction is embedded in the distance-dependent hopping parameter  $t_{n+1,n}$.
In the first-order approximation, 
$t_{n+1,n}$ linearly depends on the relative atomic displacement:
$
t_{n+1,n} = t_0 - \alpha (u_{n+1} - u_{n}),
$
where $\alpha$ and $t_0$ are set to be positive for $p$-orbital systems.

Both models undergo a spontaneous Peierls dimerization~\cite{gruner1988},
which gives two (energetically degenerate) dimerized ground states as shown in Figs.~\ref{fig:models}(b) and \ref{fig:models}(d);
$\Delta > 0$ and $\Delta < 0$ correspond to the $A$ ($A^*$) and $B$ ($B^*$) phases for the SSH (RM) model, respectively.
These two ground states are distinguished by the dimerization displacement $u_n = (-1)^{n+1} u$, where $u>0$ and $u<0$ correspond to $A$ ($A^*$) and $B$  ($B^*$) phases.
Here $u$ (or, equivalently, $\Delta = 4 \alpha u$) is the dimerization displacement of the ground states.

From the tight-binding Hamiltonian, the Fourier transformed Hamiltonian is given by
\begin{eqnarray} 
H_{\text{single}} = \sum_{k_x, s}  \sum_{i, j=1}^{2} \mathcal{H}^{ij}_{\text{single}}(k_x, \Delta, m_z) c^{\dagger}_{i, k_x, s} c_{j, k_x, s},
\end{eqnarray}
where the single-chain Bloch Hamiltonian $\mathcal{H}_{\text{single}} (k_x, \Delta, m_z ) $ is given by
\begin{equation}\label{eq:single_Bloch_Hamiltonian}
\mathcal{H}_{\text{single}} = 2t_0 \cos (\tfrac{k_x a_0}{2}) \sigma_x  -\Delta  \sin (\tfrac{k_x a_0}{2})  \sigma_y +  m_z \sigma_z.
\end{equation}
Here $\sigma_i$ is the Pauli matrix for the pseudospin space and the spin index $s$ is omitted for simplicity.
The energy eigenvalues $E$ are analytically obtained as
\begin{equation} \label{eq:RM_Bloch_Hamiltonian_eigenvalue}
E  = \pm \left [ 4t_0 ^2 \cos ^2 (\tfrac{k_x a_0}{2}) + \Delta^2 \sin^2 (\tfrac{k_x a_0}{2} )+ m_z^2  \right ]^{1/2}.
\end{equation}

If one takes the continuum limit ($a_0 \rightarrow 0$) and treats the dimerization displacement as a classical field $\Delta(x)$,
then the low-energy effective continuum Hamiltonian near the Fermi level can be obtained~\cite{brazovskii1980, takayama1980, RM1982, jackiw1983}.
Then the low-energy effective Hamiltonian $\mathsf{H}_{\text{single}} (x, \Delta(x), m_z)$ is given by
\begin{eqnarray}
\mathsf{H}_{\text{single}}  =  - i v_F \partial_x \sigma_x  + \Delta(x) \sigma_y + m_z \sigma_z,
\end{eqnarray}
where $v_{F} = t_0 a_0$ is the Fermi velocity.

\subsection{Double-chain model}
The DC model---which is a nontrivially extended model---is composed of two identical SSH chains without (with) a zigzag interchain coupling $\delta$
as shown in  Fig.~\ref{fig:models}(e) [Fig.~\ref{fig:models}(f)].
The DC model is described by 
the tight-binding Hamiltonian $H_{\text{DC}}$
which is given by
\begin{eqnarray}
H_{ \text{DC} } = 
H^{(1)}_{ \text{SSH} } + H^{(2)}_{ \text{SSH} } 
+ H_{\text{coupling}},
\end{eqnarray}
where 
\begin{eqnarray*}
H^{(i)}_{ \text{SSH} } &=& H^{(i)}_{e} + H^{(i)}_{\text{ph}}, 
\\
H^{(i)}_{e}  &=& \sum_{n, s} t^{(i)}_{n+1, n} {c^{(i)} }^{\dagger}_{n+1, s} c^{(i)}_{n, s}  + \text{H.c.},
\\
H^{(i)}_{ \text{ph} } &=&  \sum_{n=1}^{N} \frac{1}{2} M ( \dot{ u}_{n}^{(i)})^2
+ \frac{1}{2} K \left[ u^{(i)}_{n+1} - u^{(i)}_{n}  \right]^2, 
\end{eqnarray*}
and
\begin{eqnarray*}
&& H_{\text{coupling}} = \delta  t_0 \sum_{n, s } 
 \left [ {c^{(1)}_{n, s}}^{\dagger}   c^{(2)}_{n, s} 
  + {c^{(1)}_{n, s}}^{\dagger}   c^{(2)}_{n+1, s}  + \text{H.c.} \right ].
\end{eqnarray*} 
Here the superscript $i$ indicates the $i$th chain ($i=1, 2$)
and $\delta$ indicates the interchain coupling strength [zigzag dashed lines in Fig.~\ref{fig:models}(f)] between the two chains.
The interchain coupling is assumed to be a constant regardless of the atomic distance for simplicity.

\begin{figure}[tb]
\centering \includegraphics[width = 80mm]{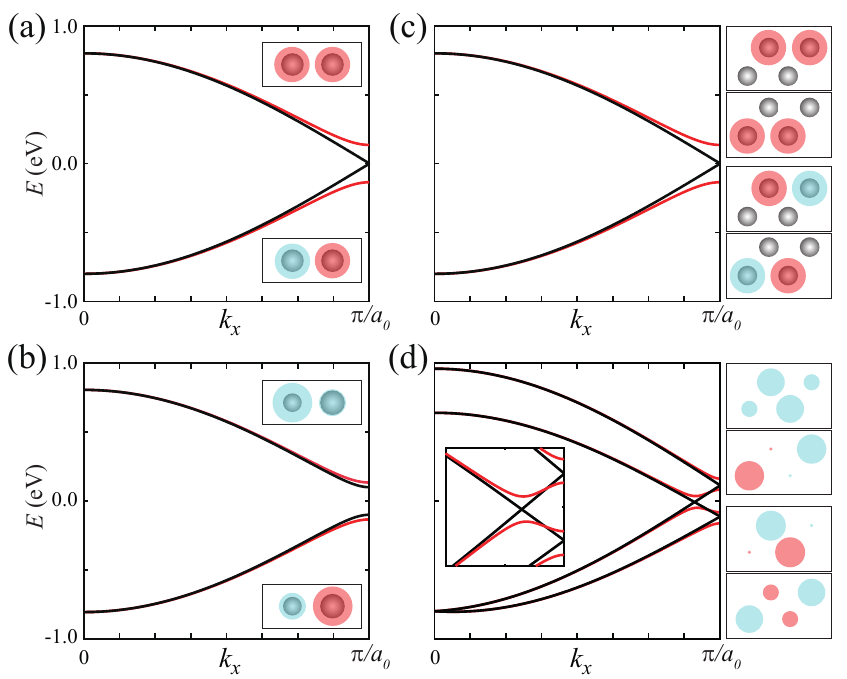}
\caption{ \label{fig:band}
[(a)--(d)] Band structures of the (a) SSH, (b) RM, and DC models (c) without and (d) with interchain coupling.
The red and black bands represent the bands of the dimerized and undimerized phases in each model, respectively.
In (d), the left inset shows the closeup from $\frac{3}{4}\frac{\pi}{a_0}$ to $\frac{\pi}{a_0}$.
The band structures for the SSH and RM models in (a) and (b)
have the spectral symmetry---particle-hole symmetric spectra with respect to $E=0$.
For the DC model with an interchain coupling, the band structure near the Fermi level [left inset of (d)] shows the spectral symmetry.
The wave functions of the dimerized phases at $k_x = \pi / a_0$
are plotted in each inset.
The amplitude and phase of the wave function are represented by the size of circles and colors, respectively.
Red and cyan colors indicate $0$ and $\pi$ phases, respectively.
In each inset,
the atoms in a unit cell are shown while they are omitted for simplicity in the right inset of (d).
}
\end{figure}

Similarly to the single-chain models, 
the DC model undergoes the spontaneous Peierls dimerization.
Instead of the two degenerate phases in the single-chain model,
the DC model has four energetically degenerate phases [$AA$, $AB$, $BA$, and $BB$ in Fig.~\ref{fig:band}(d)].
The four phases are distinguished by dimerization displacements of each chain,
$u_n^{(i)}  = (-1)^{n+1} u^{(i)}$,
where $u^{(i)}>0$ and $u^{(i)}<0$ correspond to $A$ and $B$ phases for the $i$th chain, respectively.
$u^{(i)}$ [or, equivalently, $\Delta^{(i)} = 4 \alpha u^{(i)}$]
is the dimerization displacement of the $i$th chain.

Then the Fourier transformed Hamiltonian is given by
\begin{eqnarray}\label{eq:Bloch_Ham_double_chain}
H_{\text{DC}}
= \sum_{k_x, s} \sum_{i,j=1}^{4} \mathcal{H}_{\text{DC}}^{ij}
 (k_x, \Delta^{(1)},\Delta^{(2)}) c^{\dagger}_{i, k_x, s} c_{j, k_x, s}, 
\end{eqnarray}
where the Bloch Hamiltonian is given by
\begin{eqnarray}\label{eq:Bloch_Double_Chain_Hamiltonian}
&& \mathcal{H}_{\text{DC}} (k_x, \Delta^{(1)},\Delta^{(2)})  =
\begin{pmatrix}
\mathcal{H}_{1} & \mathcal{H}_{12}\\
\mathcal{H}_{21} & \mathcal{H}_{2}
\end{pmatrix},
\end{eqnarray}
with
\begin{eqnarray*}
&& \mathcal{H}_{i}   = ( 2t_0 \cos ( k_x a_0/2 ) , -\Delta^{(i)} \sin ( k_x a_0 / 2 ), 0 ) \cdot \boldsymbol{\sigma},
\\
&& \mathcal{H}_{12} = \mathcal{H}^{\dagger}_{21} = \delta t_0 (e^{- i k_x a_0/4} 1_{2 \times 2} + e^{i k_x a_0/4} \sigma_x).
\end{eqnarray*}

Similarly to the single-chain model, the low-energy effective continuum Hamiltonian near the Fermi level is given by
\begin{eqnarray}\label{eq:Continuum_Double_Chain_Hamiltonian}
&&\mathsf{H}_{\text{DC}} (\Delta^{(1)},\Delta^{(2)})  =
\begin{pmatrix}
\mathsf{H}_{1} & \mathsf{H}_{12}\\
\mathsf{H}_{21} & \mathsf{H}_{2}
\end{pmatrix},
\end{eqnarray}
where
\begin{eqnarray}
&& \mathsf{H}_{i} =  - i v_F \partial_x \sigma_x  + \Delta^{(i)}(x) \sigma_y, \\
&& \mathsf{H}_{12} = \mathsf{H}_{21}^{\dagger}
 =   \frac{t_0 \delta}{ \sqrt  2 } \left[ (1+i) 1_{2 \times 2}  + (1-i) \sigma_x  \right].
\end{eqnarray}
All four ground states have the same energy eigenvalues, which are given by
\begin{eqnarray} \label{eq:double_continuum_eigenvalues}
E_{k_x}  = \pm \sqrt {A \pm B}, 
\end{eqnarray}
where $ A= v_F^2 k_x^2 + \Delta^2+ 2(t_0 \delta )^2, B = 2 t_0 \delta \sqrt { 2 v_F^2 k_x^2 + \Delta^2 }$.
%

\subsection{Band structure and sublattice symmetry}
Before going on, we briefly discuss the band structures and the sublattice symmetries.
For the SSH and RM models, the calculated band structures [Figs.~\ref{fig:band}(a) and \ref{fig:band}(b)] indicate that both SSH and RM models have the spectral symmetry---particle-hole  symmetric  spectra.
This particle-hole symmetry can be seen in the energy eigenvalues
in Eq.~(\ref{eq:RM_Bloch_Hamiltonian_eigenvalue}), which will be consistently explained by the symmetry analysis in Sec.~\ref{sec:symmetry_analysis}.
The calculated wave functions [insets in Figs.~\ref{fig:band}(a) and \ref{fig:band}(b)] show sublattice symmetry in the SSH model while they do not in the RM model.

For the DC model, in the absence of the interchain coupling  ($\delta = 0$),
the band structure is the duplication of the SSH bands [Fig.~\ref{fig:band}(c)].
In the presence of the interchain coupling  ($\delta \neq 0$),
the band structure does not have a spectral symmetry in the whole Brillouin zone  
due to dynamical sublattice symmetry breaking [Fig.~\ref{fig:band}(d)].
Near the Fermi level, however, there exists a spectral symmetry [see left inset of Fig.~\ref{fig:band}(d)].
This spectral symmetry can be explained by the energy eigenvalues
in Eq.~(\ref{eq:double_continuum_eigenvalues}) in the low-energy effective theory, which will be consistently explained by the symmetry analysis in Sec.~\ref{sec:symmetry_analysis}.

\section{Symmetries of Hamiltonians} \label{sec:symmetry_analysis}
We briefly discuss the limitations of three fundamental nonspatial symmetry operators ($\hat{\mathcal{T}}, \hat{\mathcal{C}}$, and $\hat \Gamma$).
For instance, in the general single-chain model, these operators are represented in the Bloch basis as
\begin{eqnarray}
\hat{\mathcal{T}} & = & \hat{ K } \otimes ( k_x \rightarrow -k_x ) ,
\\
\hat{ \mathcal {C}} & = & \sigma_z \hat{K} \otimes (k_x\rightarrow-k_x),
\\
\hat{ \Gamma } & =  &   \sigma_z,
\end{eqnarray}
and they satisfy the \textit{prior} relation~\cite{chiu2016} of $\hat{ \Gamma }= \hat{\mathcal{T}} \hat{ \mathcal {C}} $.
Here $\hat{K}$ is the complex conjugation operator.
Under these operations, the general single-chain Bloch Hamiltonian 
in Eq.~(\ref{eq:single_Bloch_Hamiltonian}) satisfies the following equations:
\begin{eqnarray*}
\hat{\mathcal{T}} \mathcal{H}_{\text{single}} (k_x, \Delta, m_z ) \hat{\mathcal{T}} ^{ -1 } 
& = & + \mathcal{H}_{\text{single}} (k_x, \Delta, +m_z ),
\\
\hat{ \mathcal {C}} \mathcal{H}_{\text{single}} (k_x, \Delta, m_z ) \hat{ \mathcal {C}}^{ -1 } 
& = & - \mathcal{H}_{\text{single}} (k_x, \Delta, -m_z ),
\\
\hat{ \Gamma } \mathcal{H}_{\text{single}} (k_x, \Delta, m_z ) \hat{ \Gamma }^{ -1 } 
& = & - \mathcal{H}_{\text{single}} (k_x, \Delta, -m_z ).
\end{eqnarray*}
Thus, the SSH model has time-reversal, particle-hole, and chiral symmetries while the RM model has time-reversal symmetry only.

For the DC model, the time-reversal $\hat{\mathcal{T}}_{\text{D}}$,  particle-hole $\hat{ \mathcal{C} }_{\text{D}}$, and  chiral  $\hat{\Gamma}_{\text{D}}$ symmetry operators are given by the direct sum of the corresponding operators for each chain.
The explicit form of each operators are shown in Table \ref{appendix:table:basic_operators_double_chain_model}
in Appendix~\ref{appendix:classI_II_IIIoperators}.
Then the DC Bloch Hamiltonian in Eq.~(\ref{eq:Bloch_Double_Chain_Hamiltonian}) has the time-reversal symmetry while not having the particle-hole and chiral symmetries in the presence of the interchain coupling:
\begin{eqnarray*}
\hat{ \mathcal{T} }_{\text{D}} \mathcal{H}_{\text{DC}} (k_x, \Delta^{(1)}, \Delta^{(2)} ) \hat{ \mathcal{T} }^{-1}_{\text{D}}
	& = &   + \mathcal{H}_{\text{DC}} (k_x, \Delta^{(1)}, \Delta^{(2)}),
	\\
\hat{\mathcal{C}}_{\text{D}} \mathcal{H}_{\text{DC}} (k_x, \Delta^{(1)},\Delta^{(2)}) ~ \hat{\mathcal{C}}^{-1}_{\text{D}}
	& \neq &  -  \mathcal{H}_{\text{DC}} (k_x, \Delta^{(1)},\Delta^{(2)}),
	\\
\hat{\Gamma}_{\text{D}} \mathcal{H}_{\text{DC}} (k_x, \Delta^{(1)},\Delta^{(2)}) ~ \hat{\Gamma}^{-1}_{\text{D}}
	& \neq &  -  \mathcal{H}_{\text{DC}} (k_x, \Delta^{(1)},\Delta^{(2)}).
\end{eqnarray*} 

Therefore, SSH model is in the BDI class while the RM and DC ones are in the AI class (Altland-Zirnbauer classification~\cite{Altland1997,schnyder2008}).
As a result, $\hat{\mathcal{T}}, \hat{\mathcal{C}}$, and $\hat \Gamma$ operators are not sufficient to discuss the properties of the ground states and topological solitons for all three systems in a single framework.
%
%

\begin{table*}[ht]
\centering
\caption{
\label{table:transformation_hamiltonian} 
Transformation properties of  the Hamiltonians of the SSH, RM, and DC model under the class~I, II, and III operators.
A transformed Hamiltonian  under an operation $\hat{\mathcal{O}}$ is given by the following transformation equation:
either $\hat{\mathcal{O}}  H (k, \Delta, m_z) \hat{\mathcal{O}}^{-1} = \eta H (k, \Delta\rq{} , m_{z}') $ for the SSH and RM models or
$\hat{\mathcal{O}} H (k, \Delta^{(1)},\Delta^{(2)} ) \hat{\mathcal{O}}^{-1} = \eta H (k, \Delta\rq{}^{(1)},\Delta\rq{}^{(2)} ) $ for the DC model, where $\eta = \pm 1$.
For the SSH and RM models, 
$H$ can be the Bloch and low-energy effective Hamiltonians
to all class~I, II, and III operators.
For the DC model,
$H$  can be the Bloch and low-energy effective Hamiltonians
to the class~I operator
while $H$ is the low-energy effective Hamiltonian to the class~II and III operators.
Here $X$, $Y$ and $Z$ can be either $A$ or $B$ phases for the SSH and DC models; either $A^*$ or $B^*$ phases for the RM model.
$X$ and $Y$ cannot be the same.
%
``GS connecting'' means that the corresponding operator connects different ground states
and ``Chiral symmetry'' means that the corresponding operator endows the chiral symmetry to a ground state itself.
Under the column  ``Group'', 
$\mathbb{Z}_n$ indicates that the corresponding operator connects $n$ different ground-state phases cyclically.
}
{\renewcommand{\arraystretch}{0.6}%
\begin{tabular}{@{\extracolsep{20pt}} l c c c c c c c c}
\hline
\hline
\\
Model & Class & Operator & ~~~$\eta$~~~  & \multicolumn{2}{c}{~~~~$\Delta'$s \& $m_z'$}
& ~~~~~~ GS relation ~~~~~~ & ~~~~~~~ Role ~~~~~~~ & Group \\ 
\\
\hline
\\
& I & $ \hat{\mathcal{O}}_{\text{I}} $
& $+1$ & $-\Delta$ & 0 &$ \hat{\mathcal{O}}_{\text{I}} X = Y$
& GS connecting & $\mathbb{Z}_2$ 
\\
SSH
&II & $\hat{\Gamma}_{\text{II}}$, $\hat{ \mathcal{C}}_{\text{II}} $
& $-1$ & $-\Delta$ & 0 & $\hat{\mathcal{O}}_{\text{II}}X=Y$
& GS connecting & $\mathbb{Z}_2$ 
\\
&III & $\hat{\Gamma}_{\text{III}}$, $\hat{ \mathcal{C}}_{\text{III}} $
& $-1$ & $+\Delta$ & 0 & $\hat{\mathcal{O}}_{\text{III}}X=X$
& Chiral symmetry & $\mathbb{Z}_1$ 
\\
\\
\hline
\\
& I & $ \hat{\mathcal{O}}_{\text{I}} $
& $+1$ & $-\Delta$ & $+m_z$ 
&$ \hat{\mathcal{O}}_{\text{I}}X=Y$
& GS connecting & $\mathbb{Z}_2$ 
\\
RM
&II & $\hat{\Gamma}_{\text{II}}$, $\hat{ \mathcal{C}}_{\text{II}} $
& $-1$ & $-\Delta$ & $+m_z$ & $\hat{\mathcal{O}}_{\text{II}}X=Y$
& GS connecting & $\mathbb{Z}_2$ 
\\
&III & $\hat{\Gamma}_{\text{III}}$, $\hat{ \mathcal{C}}_{\text{III}} $
& $-1$ & $+\Delta$ & $+m_z$ & $\hat{\mathcal{O}}_{\text{III}}X=X$
& Chiral symmetry & $\mathbb{Z}_1$ 
\\
\\
\hline
\\
& I & $ \hat{\mathcal{O}}_{\text{I}} $
& $ +1 $ & ~~~~$ -\Delta^{(2)} $ & \hspace{5pt}$ + \Delta^{(1)} $ &
See below\footnote{$AA = \hat{\mathcal{O}}_{\text{I}} AB = \hat{\mathcal{O}}_{\text{I}}^2 BB =\hat{\mathcal{O}}_{\text{I}}^3 BA =\hat{\mathcal{O}}_{\text{I}}^4 AA$}
& GS connecting & $\mathbb{Z}_4$ 
\\
& II & $ \hat{\Gamma}_{\text{II}}^{(1)}$,  $\hat{\mathcal{C}}^{(1)}_{\text{II}} $
& $ -1 $ & ~~~~$ -\Delta^{(1)} $ & \hspace{5pt}$ + \Delta^{(2)} $ & $\hat{\mathcal{O}}^{(1)}_{\text{II}} XZ = YZ $
& GS connecting & $\mathbb{Z}_2$ 
\\
DC
& II & $\hat{\Gamma}_{\text{II}}^{(2)}$, $\hat{ \mathcal{C}}^{(2)}_{\text{II}} $
& $ -1 $ & ~~~~$ + \Delta^{(1)} $ & \hspace{5pt}$ -\Delta^{(2)} $ & $\hat{\mathcal{O}}^{(2)}_{\text{II}} ZX = ZY $
& GS connecting & $\mathbb{Z}_2$ 
\\
& III & $ \hat{\Gamma}_{\text{III}}^{(1)}$, $ \hat{\mathcal{C}}_{\text{III}}^{(1)}$
& $ -1 $ & ~~~~$ + \Delta^{(2)} $ & \hspace{5pt}$ + \Delta^{(1)} $ 
& $ \hat{\mathcal{O}}_{\text{III}}^{(1)} XX = XX $
& Chiral symmetry & $\mathbb{Z}_1$ 
\\
& & 
& &  &  
& $ \hat{\mathcal{O}}_{\text{III}}^{(1)} XY = YX $
& GS connecting & $\mathbb{Z}_2$ 
\\
& III & $ \hat{\Gamma}_{\text{III}}^{(2)}$, $ \hat{\mathcal{C}}_{\text{III}}^{(2)}$
& $ -1 $ & ~~~~$ -\Delta^{(2)} $ & \hspace{5pt}$ -\Delta^{(1)} $ 
& $ \hat{\mathcal{O}}_{\text{III}}^{(2)} XY = XY  $
& Chiral symmetry &  $\mathbb{Z}_1$
\\
& & 
& &  &  
& $ \hat{\mathcal{O}}_{\text{III}}^{(2)} XX = YY $
& GS connecting & $\mathbb{Z}_2$ 
\\
\\
\hline \hline
\end{tabular}}

\end{table*}

To overcome this limitation, we construct class~I, II, and III operators using the three nonspatial symmetry operators above and crystalline nonsymmorphic symmetry operators. 
The class~I operators ($\hat{\mathcal{O}}_{\text{I}}$) are nonsymmorphic operators. For class~II operators ($\hat{\mathcal{O}}_{\text{II}}$), there are two types of operators:
a nonsymmorphic particle-hole operator ($\hat{\mathcal{C}}_{\text{II}}$) and a nonsymmorphic chiral operator ($\hat{\Gamma}_{\text{II}}$).
These operators satisfy
$\hat{\Gamma}_{\text{II}} = \hat{\mathcal{T}} \hat{\mathcal{C}}_{\text{II}} $,
similarly to $\hat \Gamma = \hat{\mathcal{T}} \hat{\mathcal{C}}$.
By multiplying class~I and II operators, the two types of class~III operators ($\hat{\mathcal{O}}_{\text{III}}$) are naturally defined:
$ \hat{\mathcal{C}}_{\text{III}} $ and  $ \hat{\Gamma}_{\text{III}} $.
Once the class~I, II, and III operators are defined,
three models have similar symmetry properties in the low-energy effective theory.
For undimerized phases, class~I, II, and III operators are symmetry operators while class~I and II are not a symmetry operators for dimerized phases as shown in Table~\ref{appendix:table:CTP_extended}.

We find that class~I, II, and III operators endow unique properties among the ground states regardless of the model,
which are summarized in Table \ref{table:transformation_hamiltonian}.
The class~I and II operators are symmetry operators before the Peierls dimerization.
After spontaneous dimerization, on the other hand, the class~I (II) operator connects distinct ground states with the same (opposite) energy spectra.
Thus, the class~I (II) operator gives the equivalent relation (particle-hole duality) among distinct ground states.
By contrast, the class~III operator acts as the generalized particle-hole and chiral operator, endowing particle-hole or chiral symmetries to the Hamiltonian.
As a consequence, $\hat{\mathcal{T}}$,
$ \hat{\mathcal{C}}_{\text{III}} $, and  $ \hat{\Gamma}_{\text{III}} $
act as the generalized $\hat{\mathcal{T}}, \hat{\mathcal{C}}$, and $\hat{\Gamma}$ symmetry operators for various quasi-1D systems as $\hat{\mathcal{T}}, \hat{\mathcal{C}}$, and $\hat{\Gamma}$ do in the SSH model.
Note that the explicit representations of the class~I, II, and III operators for SSH, RM, and DC models are shown in Appendix \ref{appendix:classI_II_IIIoperators}
(see Tables~\ref{appendix:table:operators_single_chain_model} and \ref{appendix:table:operators_double_chain_model}). 
%

\begin{figure}[b]
\centering \includegraphics[width=86mm]{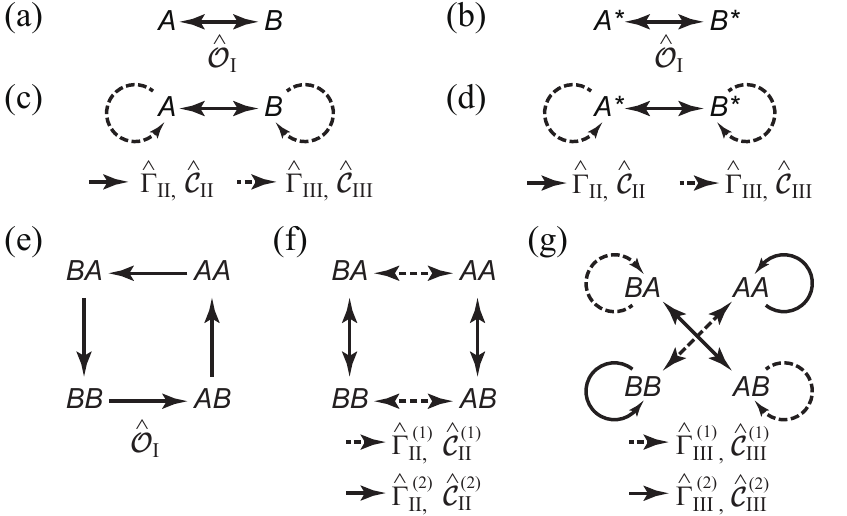} 
\caption{ \label{fig:groundclass2,3}
[(a)--(d)] Transformation properties between two degenerate ground states of the [(a) and (c)] SSH and [(b) and (d)] RM under class~I, II, and III operators. 
[(e)--(g)] Transformation properties among four degenerate ground states of the DC model under class~I, II, and III operators. 
}
\end{figure}

From now on, we will discuss the main results.
(The results are summarized in Table~\ref{table:transformation_hamiltonian}
and the corresponding schematic diagrams are shown in Fig.~\ref{fig:groundclass2,3}.)
In the SSH model, the nonsymmorphic half-translation operator, $\hat{G} \equiv \{ E | \frac{a_0}{2} \}$ is the class~I operator that connects two degenerate ground states
[Figs.~\ref{fig:models}(b) and \ref{fig:groundclass2,3}(a)]:
Mathematically,
$ \hat{G}\mathcal{H}_{\text{single}} ( k_x , \Delta , m_z=0)   \hat{G} ^{-1}  = \mathcal{H}_{\text{single}} ( k_x , - \Delta , m_z=0)$.
In the RM model, however, $\hat{G}$ does not connect two degenerate ground states due to the sublattice symmetry breaking [Fig.~\ref{fig:models}(d)] and hence it is no longer a proper class~I operator.
We find that the combined operator $\hat{G}^{\text{PT}} \equiv \hat{G} \hat{\mathcal{P}}\hat{\mathcal{T}}$
[parity-time- (PT) symmetric half-translation operator] is the class~I operator regardless of the sublattice symmetry breaking because $\hat{G}^{\text{PT}}$ shifts a half unit cell after switching $a$ and $b$ atoms [Figs.~\ref{fig:models}(d) and \ref{fig:groundclass2,3}(b)].
Here $ \hat{\mathcal{P}}$ is an inversion operator with respect to the unit-cell center.
Mathematically, 
$ \hat{G}^{\text{PT}}   \mathcal{H}_{\text{single}} ( k_x , \Delta , m_z)   (\hat{G}^{\text{PT}})^{-1}
  = \mathcal{H}_{\text{single}} ( k_x , - \Delta , m_z) $.

For the class~II operator, a nonsymmorphic chiral operator $\hat{ \Gamma }_{\text{II}} \equiv \{ E | \frac{a_0}{2} \} \otimes \hat{\Gamma} $
consistently connects two degenerate ground states in the SSH and RM models
[Figs.~\ref{fig:groundclass2,3}(c) and \ref{fig:groundclass2,3}(d)]
because $\hat{ \Gamma }$ 
exchanges the sublattice potential.
Furthermore, there exists another class~II operator, a nonsymmorphic charge-conjugation operator $ \hat{ \mathcal {C} }_{\text{II}} \equiv  \{ E | \frac{a_0}{2} \} \otimes  \hat{\mathcal{C}} $ using the relation $ \hat{\Gamma} = \hat{ \mathcal{T} } \hat { \mathcal{C} }$.
Mathematically, 
$\hat{ \Gamma }_{\text{II}} \mathcal{H}_{\text{single}} ( k_x , \Delta , m_z)  \hat{ \Gamma }_{\text{II}}^{-1} = - \mathcal{H}_{\text{single}} ( k_x , - \Delta , m_z)$
and 
$ \hat{ \mathcal {C}}_{\text{II}}  \mathcal{H}_{\text{single}} ( k_x , \Delta , m_z)   \hat{ \mathcal {C}}_{\text{II}} ^{-1}
  = - \mathcal{H}_{\text{single}} ( k_x , - \Delta , m_z) $.
Thus, class~II operators connect distinct ground states having the opposite dimerization and energy spectra leading to the particle-hole duality between ground states.

By combining the class~I and II operators, we find that 
$\hat{\mathcal{C}}$ and $\hat{\Gamma}$ ($\hat{\mathcal{C}}^{\text{PT}} $ and $\hat{\Gamma}^{\text{PT}} $) are the class~III operators for the SSH (RM) model [Figs.~\ref{fig:groundclass2,3}(c) and \ref{fig:groundclass2,3}(d)].
Here $\hat{\mathcal{C}}^{\text{PT}} \equiv  \hat{\mathcal{C}} \hat {\mathcal{P}} \hat {\mathcal{T}}$ and $\hat{\Gamma}^{\text{PT}} \equiv  \hat {\Gamma} \hat {\mathcal{P}} \hat {\mathcal{T}}$ are PT-symmetric particle-hole and chiral operators, respectively.
These class~III operators ($ \hat{ \mathcal {O}}_{\text{III}} $)  anticommute with the Hamiltonian, $ \{ \hat{ \mathcal {O}}_{\text{III}} , \mathcal{H}_{\text{single}}  \}  = 0  $, explaining the spectral symmetry of the band structures in Figs.~\ref{fig:band}(a) and \ref{fig:band}(b) regardless of the sublattice symmetry breaking.
%
%

In the DC model, as a class~I operator, the glide reflection operator $\hat{G}_y=  \{ M_y | -\frac{a_0}{4} \}$ connects the four degenerate ground states cyclically, where $M_y$ is a reflection operator with respect to the $xz$ plane [Figs.~\ref{fig:models}(g) and \ref{fig:groundclass2,3}(e)].
Mathematically, under $\hat{G}_y$, the Bloch Hamiltonian $\mathcal{H}_{\text{DC}}$
and the low-energy effective Hamiltonian $\textsf{H}_{\text{DC}}$ transform
as
\begin{eqnarray*} 
\hat{G}_y \mathcal{H}_{\text{DC}}( \Delta^{(1)}, \Delta^{(2)}) \hat{G}_y^{-1} &= &\mathcal{H}_{\text{DC}}( -\Delta^{(2)}, \Delta^{(1)}), \\
\hat{G}_y \textsf{H}_{\text{DC}}( \Delta^{(1)}, \Delta^{(2)}) \hat{G}_y ^{-1} &=& \textsf{H}_{\text{DC}}( -\Delta^{(2)}, \Delta^{(1)}),
\end{eqnarray*}
which imply a ground state with $( \Delta^{(1)}, \Delta^{(2)})$ transforms into another ground state with $(-\Delta^{(2)}, \Delta^{(1)})$.
Then a ground state comes back to itself under $\hat{G}_y^4$ and hence $\hat{G}_y$ transforms a ground state to another ground state cyclically:
\begin{eqnarray*} 
AA = \hat{G}_y AB = \hat{G}_y^2 BB = \hat{G}_y^3 BA = \hat{G}_y^4 AA.
\end{eqnarray*}
Thus, the four ground states correspond to the atomic representation of a cyclic $\mathbb{Z}_4$ group, of which the generator is $\hat{G}_y$.
In this sense, $ \hat{G}_{y} $ is a $\mathbb{Z}_4$ operator while $\hat{G}$ in the single-chain models is a $\mathbb{Z}_2$ one.
Hence, class~I operators can endow distinct group structures for degenerate ground states of different 1D electronic systems.
%

\begin{figure}[tb]
\centering \includegraphics[width = 86mm]{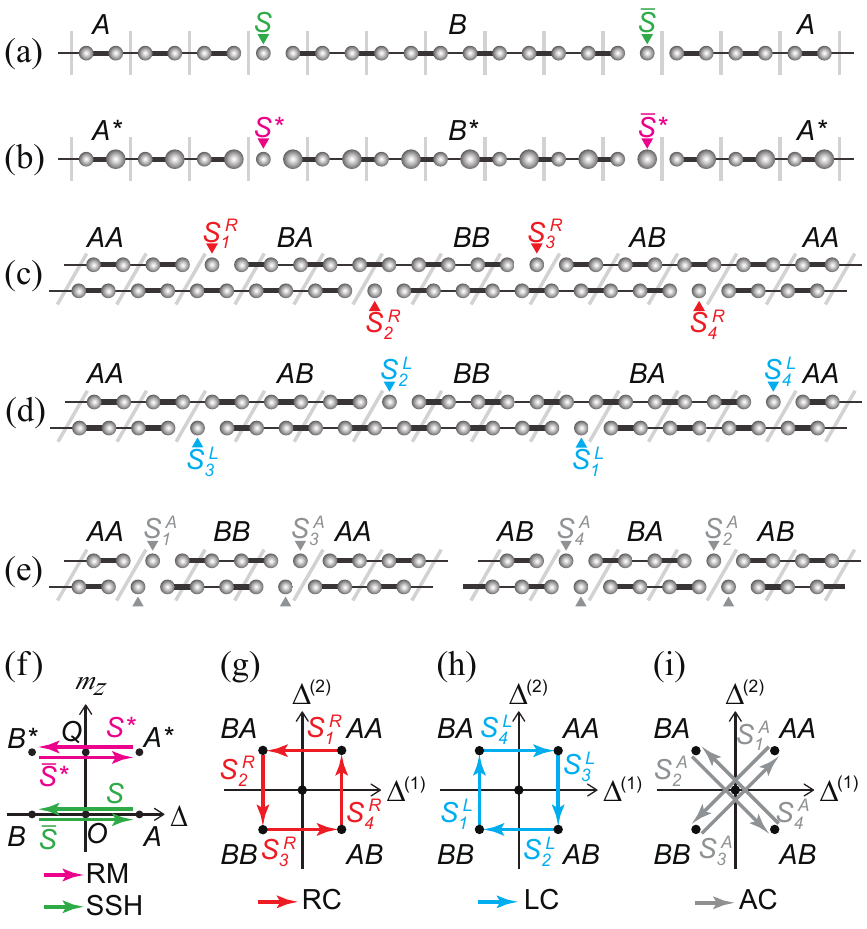}
\caption{ \label{fig:solitons} 
[(a)--(e)] Geometric configurations of solitons.
For the SSH (RM) model,
the soliton $S$ ($S^*$) and the antisoliton $\bar{S}$ ($\bar{S}^*$) are shown in (a) [(b)].
For the DC model, there are 12 chiral solitons grouped into three types:
(c) four right-chiral (RC), (d) four left-chiral (LC), and (e) four achiral (AC) solitons are shown ($S_i^k$; $i=1,2,3,4$; $k=R,L,A$).
For simplicity, all solitons are drawn as short as possible.
[(f)--(i)] Solitons in the order parameter space.
In (f), solitons of the SSH and RM models are denoted as green and magenta arrows, respectively, in the order parameter space $(\Delta, m_z)$,
where  $m_z$ is a staggered sublattice potential for the RM model.
For (g) RC, (h) LC, and (i) AC solitons, 
each soliton  is denoted by a corresponding colored arrow in the order parameter space $(\Delta^{(1)},\Delta^{(2)})$.
} 
\end{figure}

\begin{figure}[ht]
\centering \includegraphics[width=86mm]{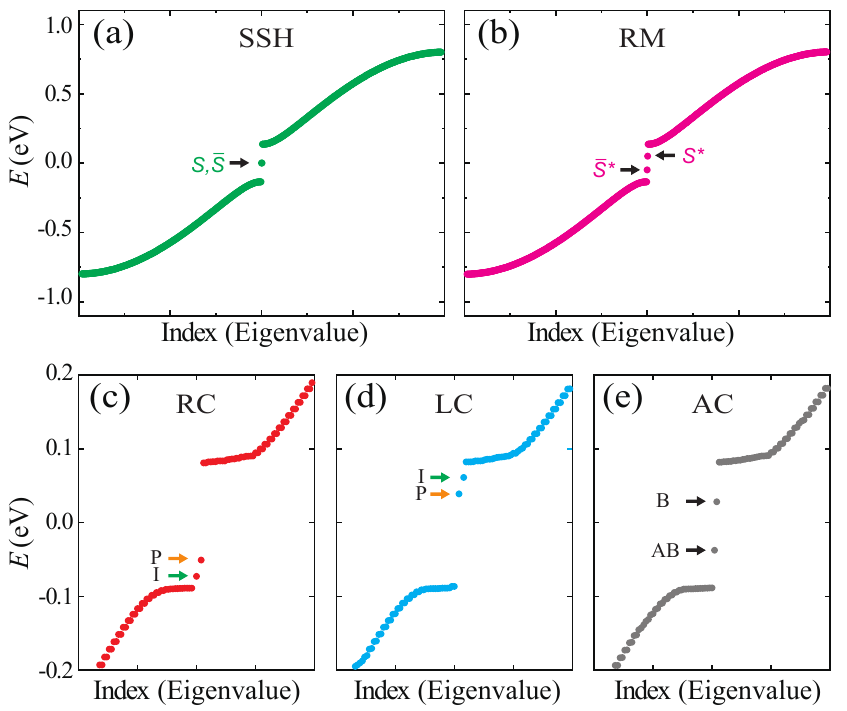} 
\caption{\label{fig:soliton_spectra}
Numerically calculated soliton spectra for the (a) SSH, (b) RM, (c) RC, (d) LC, and (e) AC solitons.
(a) In the SSH model, the soliton ($S$) and antisoliton ($\bar{S}$) states are degenerated at the midgap. 
(b) In the RM model, the soliton ($S^*$) and antisoliton ($\bar{S}^*$) states are located above ($E = +m_z$) and below ($E=-m_z$) the Fermi level, respectively. 
[(c) and (d)] RC and LC soliton has two soliton states: The orange and green arrows
indicate the primary (P) and induced (I) subsoliton states, respectively.
(e) An AC soliton has two subsoliton states which are symmetrically located with respect to the zero energy in contrast to RC and LC solitons:
the antibonding (AB) and bonding (B) subsoliton states.
In (a)--(e), all spectra are plotted as a function of the eigenvalue index.
} 
\end{figure}

The DC model has more abundant symmetry operators because it is intrinsically an extended system.
As the class~II operators, there exist two nonsymmorphic chiral operators:
$ \hat{\Gamma}^{(i)}_{\text{II}} \equiv 
\hat{G}^{(i)} \otimes \hat{\Gamma}_{\text{D}} $,
where $\hat{\Gamma}_{\text{D}} = \sigma_z \oplus \sigma_z$ is an extended chiral operator for the DC model
and $ \hat{G}^{(i)} \equiv \{ E | (-1)^{i+1} \frac{a_0}{2} \}^{(i)} $ is a half-translation operator for the $i$th chain only.
Thus, 
$\hat{\Gamma}^{(i)}_{\text{II}}$ properly connects two ground states having opposite energy spectra and dimerization pattern of the $i$th chain, leading  to  the  particle-hole  duality  between degenerate ground states [Fig.~\ref{fig:groundclass2,3}(f)].
Similarly to the single-chain model, two class~II nonsymmorphic particle-hole operators $\hat{\mathcal{C}}^{(i)}_{\text{II}}$ can be defined using $ \hat{\Gamma} = \hat{ \mathcal{T} } \hat { \mathcal{C} }$.
The mathematical proof is as follows.
Under $ \hat{\Gamma}^{(i)}_{\text{II}}$ and $\hat{ \mathcal{C} }^{(i)}_{\text{II}}$,
the low-energy effective Hamiltonian
satisfies the following transformation equations:
\begin{eqnarray*}
\hat{\Gamma}^{(1)}_{\text{II}}  \mathsf{H}_{\text{DC}} (\Delta^{(1)}, \Delta^{(2)} ) (\hat{\Gamma}^{(1)}_{\text{II}})^{-1}
 &=&   - \mathsf{H}_{\text{DC}} (-\Delta^{(1)}, \Delta^{(2)} ) , \\
\hat{\Gamma}^{(2)}_{\text{II}}  \mathsf{H}_{\text{DC}} (\Delta^{(1)}, \Delta^{(2)} ) (\hat{\Gamma}^{(2)}_{\text{II}})^{-1}
 &=&   - \mathsf{H}_{\text{DC}} (\Delta^{(1)}, -\Delta^{(2)} ) , \\
\hat{\mathcal{C}}^{(1)}_{\text{II}} \mathsf{H}_{\text{DC}} (\Delta^{(1)}, \Delta^{(2)} ) (\hat{\mathcal{C}}^{(1)}_{\text{II}})^{-1}
 &=&   - \mathsf{H}_{\text{DC}} (-\Delta^{(1)}, \Delta^{(2)} ), \\
\hat{\mathcal{C}}^{(2)}_{\text{II}} \mathsf{H}_{\text{DC}} (\Delta^{(1)}, \Delta^{(2)} ) (\hat{\mathcal{C}}^{(2)}_{\text{II}})^{-1
} &=&   - \mathsf{H}_{\text{DC}} (\Delta^{(1)}, -\Delta^{(2)} ),
\end{eqnarray*}
which imply that 
the dimerization pattern of $i$th chain and energy eigenvalues are reversed.
A ground state having energy eigenvalues $\{ \pm E_1, \pm E_2, \ldots \}$ transforms to another ground state having the same energy eigenvalues except for the energy band crossing.
Thus, class~II operators connect different phases as summarized in Table~\ref{table:transformation_hamiltonian}.

Finally, by combining the class~I and II  operators,
we find that the class~III operators $ \hat{\Gamma }^{(i)}_{\text{III}}$ and $ \hat{\mathcal{C} }^{(i)}_{\text{III}}$ give not only the spectral symmetry but also particle-hole dualities among the ground states [Fig.~\ref{fig:groundclass2,3}(g)].
Under $\hat{\Gamma}_{\text{III}}^{(i)} $ and $\hat {\mathcal{C}}^{(i)}_{\text{III}} $,
the low-energy effective Hamiltonian satisfies the following transformation equations:
\begin{eqnarray*}
\hat{\Gamma}_{\text{III}}^{(1)}  \mathsf{H}_{\text{DC}} (\Delta^{(1)}, \Delta^{(2)} ) ( \hat{\Gamma}_{\text{III}}^{(1)} )^{-1}
& = &  - \mathsf{H}_{\text{DC}} (+\Delta^{(2)}, +\Delta^{(1)} ) , 
\\
\hat\Gamma^{(2)}_{\text{III}}   \mathsf{H}_{\text{DC}} (\Delta^{(1)}, \Delta^{(2)} ) (\hat\Gamma^{(2)}_{\text{III}} )^{-1}
& = &  -\mathsf{H}_{\text{DC}} (-\Delta^{(2)}, -\Delta^{(1)} ),
\\
\hat {\mathcal{C}}^{(1)}_{\text{III}}  \mathsf{H}_{\text{DC}} (\Delta^{(1)}, \Delta^{(2)} ) ( \hat{\mathcal{C}}^{(1)}_{\text{III}})^{-1}
& = &  - \mathsf{H}_{\text{DC}} (+\Delta^{(2)}, +\Delta^{(1)} ) , 
\\
\hat{\mathcal{C}}^{(2)}_{\text{III}}  \mathsf{H}_{\text{DC}} (\Delta^{(1)}, \Delta^{(2)} ) (\hat{\mathcal{C}}^{(2)}_{\text{III}} )^{-1}
& = &  -\mathsf{H}_{\text{DC}} (-\Delta^{(2)}, -\Delta^{(1)} ).
\end{eqnarray*} 
The first and third equations show the chiral symmetry for $AA$ and $BB$ phases [$\Delta^{(1)}= \Delta^{(2)}$].
The second and fourth equations also indicate the chiral symmetry for $AB$ and $BA$ phases [$\Delta^{(1)}=- \Delta^{(2)}$].
Therefore, $\hat \Gamma^{(i)}_{\text{III}} $ and $\hat {\mathcal{C}}^{(i)}_{\text{III}}$
act as the generalized chiral symmetry and particle-hole operators, respectively.
Moreover, the class~III operators in the DC model also act as ground state-connecting operators;
the first and third equations connect the $AB$ and $BA$ phases [$\Delta^{(1)}= -\Delta^{(2)}$]
and  the second and fourth equations connect the $AA$ and $BB$ phases [$\Delta^{(1)}=\Delta^{(2)}$].
%

\begin{table*}[t]
\caption{
\label{table:transformation_soliton} 
Transformation properties of the solitons under the class~I, II, and III operators.
Before a transformation, a soliton has an energy eigenvalue $E$ and dimerization profiles
$\Delta(x)$ or
$(\Delta^{(1)}(x), \Delta^{(2)}(x) )$
while the transformed soliton has the energy eigenvalue $E\rq{}$ and dimerization profiles
$\Delta(x)'$ or $(\Delta\rq{}^{(1)}(x), \Delta\rq{}^{(2)}(x) )$.
For the RM model, we explicitly take the spatial inversion $x \rightarrow -x$ due to the inversion operator $\hat {\mathcal{P}}$.
Under the column ``Soliton relation,'' $X$ and $Y$ can represent $S$ ($S^*$) and $\bar S$ ($\bar S^*$) for the SSH (RM) model, 
but $X$ and $Y$ cannot be the same.
For the DC model, $S^k_{0,1} = S^k_{4,5}$ for all $k= R, L, A$.
The role of the operator is explicitly shown under the column  ``Role''
and is categorized under the column  ``Type'' according to the similarity with the SSH and RM models,
where ``Both'' represents the role of the operator in the DC model is similar to both the SSH and RM models.
Under the column ``Group,''
$\mathbb{Z}_n$ indicates that the corresponding operator connects $n$ different topological solitons cyclically.
}
{\renewcommand{\arraystretch}{0.6}%
\begin{tabular}{@{\extracolsep{5pt}} l c  c c c c c c c c}
\hline
\hline
\\
Model & Class &Operator & $ E\rq{} $ &
\multicolumn{2}{c}{ $\Delta \rq{}(x) $s}  &  Soliton relation  & Role & Type & Group \\ 
\\
\hline
\\
& I & $\hat{\mathcal{O}}_{\text{I}} $
&$+E$ &  \multicolumn{2}{c}{$-\Delta(x)$} &   $ \hat{\mathcal{O}}_{\text{I}}X = Y $
& Equivalence  && $\mathbb{Z}_2$
\\
SSH
&II & $\hat{\Gamma}_{\text{II}}, \hat{ \mathcal{C}}_{\text{II}} $
& $-E$ &
\multicolumn{2}{c}{$-\Delta(x)$} & $\hat{\mathcal{O}}_{\text{II}}X = Y $
& Particle-hole  && $\mathbb{Z}_2$
\\
&III & $\hat{\Gamma}_{\text{III}}$, $\hat{\mathcal{C}}_{\text{III}}$
& $-E$ &
\multicolumn{2}{c}{$+\Delta(x)$} & $\hat{\mathcal{O}}_{\text{III}}X = X $
& Self-duality && $\mathbb{Z}_1$
\\
\\
\hline
\\
& I & $\hat{\mathcal{O}}_{\text{I}}$
&  $+E$ &
\multicolumn{2}{c}{$-\Delta(-x)$ }& $\hat{\mathcal{O}}_{\text{I}}X = X $
& Equivalence  && $\mathbb{Z}_1$
\\
RM
&II & $\hat{\Gamma}_{\text{II}}, \hat{\mathcal{C}}_{\text{II}} $
& $-E$ &
\multicolumn{2}{c}{$-\Delta(x)$ }&  $\hat{\mathcal{O}}_{\text{II}}X = Y  $
& Particle-hole  && $\mathbb{Z}_2$
\\
&III & $\hat{\Gamma}_{\text{III}}$, $\hat{\mathcal{C}}_{\text{III}}$
& $-E$ &
\multicolumn{2}{c}{$+\Delta(-x)$ } & $\hat{\mathcal{O}}_{\text{III}}X = Y $
& Particle-hole  && $\mathbb{Z}_2$
\\
\\
\hline
\\
& I & $ \hat{\mathcal{O}}_{\text{I}} $
& $ +E $ &  $ -\Delta^{(2)}(x) $ & $ + \Delta^{(1)}(x) $
& See below\footnote{
$S_1^{k} = \hat{\mathcal{O}}_{\text{I}} S_4^{k} =  \hat{\mathcal{O}}_{\text{I}}^2 S_3^{k} = \hat{\mathcal{O}}_{\text{I}}^3 S_2^{k} = \hat{\mathcal{O}}_{\text{I}}^4 S_1^{k}$ for $k= R, L, A$.
}
& Equivalence  & SSH
& $\mathbb{Z}_4$
\\
& II & $ \hat{\Gamma}^{(1)}_{\text{II}}, \hat{\mathcal{C}}^{(1)}_{\text{II}} $
& $ -E $  & $ -\Delta^{(1)}(x) $ & $ + \Delta^{(2)}(x) $
& $\hat{\mathcal{O}}^{(1)}_{\text{II}} S_{i}^{R(L)} = S_{5-i}^{L(R)}$ $(i=1,2,3,4)$
& Particle-hole  & Both
& $\mathbb{Z}_2$
\\
&  & 
&  &  & 
&  $\hat{\mathcal{O}}^{(1)}_{\text{II}} S_{i}^{A} = S_{i+1}^{A}$ $(i=1, 3)$
& Particle-hole  & Both
& $\mathbb{Z}_2$
\\
DC
& II & $ \hat{\Gamma}^{(2)}_{\text{II}}, \hat{ \mathcal{C}}^{(2)}_{\text{II}} $
& $ -E $ & ~ $ + \Delta^{(1)}(x) $ & $ -\Delta^{(2)}(x) $ ~
& ~ $\hat{\mathcal{O}}^{(2)}_{\text{II}} S_{i}^{R(L)} = S_{i+(-1)^{i+1}}^{L(R)} $ $(i=1,2,3,4)  $ ~
& Particle-hole  & Both
& $\mathbb{Z}_2$
\\
&  & 
&  &  & 
&  $\hat{\mathcal{O}}^{(2)}_{\text{II}} S_{i}^{A} = S_{5-i}^{A}$ $(i=2, 4)$
& Particle-hole  & Both
& $\mathbb{Z}_2$
\\
& III & $ \hat{\Gamma}_{\text{III}}^{(1)}$, $ \hat{\mathcal{C}}_{\text{III}}^{(1)}$
& $ -E $  & $ + \Delta^{(2)}(x) $ & $ + \Delta^{(1)}(x) $ 
& $\hat{\mathcal{O}}_{\text{III}}^{(1)} S_{i}^{A} = S_{i}^{A} $ $(i=1, 3)$ 
& Self-duality & SSH
& $\mathbb{Z}_1$
\\
&  & 
&  &  & 
&  $\hat{\mathcal{O}}_{\text{III}}^{(1)} S_{i}^{R(L)} = S_{4-i}^{L(R)}$ $(i=1,2,3, 4)$
& Particle-hole  & RM
& $\mathbb{Z}_2$
\\
& III & $ \hat{\Gamma}_{\text{III}}^{(2)} $, $ \hat{\mathcal{C}}_{\text{III}}^{(2)} $
& $ -E $  & $ -\Delta^{(2)}(x) $ & $ -\Delta^{(1)}(x) $ 
& $\hat{\mathcal{O}}_{\text{III}}^{(2)}  S_{i}^{A} = S_{i}^{A} $ $(i=2, 4)$
& Self-duality & SSH
& $\mathbb{Z}_1$
\\
&  & 
&  &  & 
&  $\hat{\mathcal{O}}_{\text{III}}^{(2)} S_{i}^{R(L)} = S_{6-i}^{L(R)}$ $(i=1,2,3, 4)$
& Particle-hole  & RM
& $\mathbb{Z}_2$
\\
\\
\hline \hline
\end{tabular}}
\end{table*}
\section{Dualities between topological solitons}
We briefly introduce the topological solitons before discussing the their topological features.
All possible types of solitons can be represented in the order parameter spaces [Figs.~\ref{fig:solitons}(f)--\ref{fig:solitons}(i)].
Each soliton has a definite geometrical configuration [Figs.~\ref{fig:solitons}(a)--\ref{fig:solitons}(e)] and the characteristic spectrum [Figs.~\ref{fig:soliton_spectra}(a)--\ref{fig:soliton_spectra}(e)].
In the SSH (RM) model, there are two types of solitons:
a soliton $S$ ($S^*$) and an antisoliton $\bar S$ ($\bar S^*$).
Both soliton and antisoliton states in the SSH model are located at zero energy [Fig.~\ref{fig:soliton_spectra}(a)] due to the chiral symmetry.
For RM solitons, the energy spectrum of a soliton (an antisoliton) is above (below) $E=0$ [Fig.~\ref{fig:soliton_spectra}(b)] due to the chiral (or sublattice) symmetry breaking.

In the DC model, 
there are 12 chiral solitons,
which are classified into three classes~\cite{cheon2015}:
right-chiral (RC), left-chiral (LC), and achiral (AC) solitons [Figs.~\ref{fig:solitons}(c)--\ref{fig:solitons}(e)].
All chiral solitons have two midgap states
but the corresponding spectra are distinguishable.
For RC solitons (LC solitons), two states are located below (above) $E=0$ [Figs.~\ref{fig:soliton_spectra}(c) and \ref{fig:soliton_spectra}(d)];
the soliton states near (far from) the Fermi level are denoted as primary (induced) subsoliton states.
A pair of an RC soliton and an LC soliton forms a particle-hole symmetric spectra.
For AC solitons, a bonding (antibonding) state is located above (below) $E=0$ [Fig.~\ref{fig:soliton_spectra}(e)].
All these particle-hole symmetric spectra are explained by the symmetry analysis below.

\begin{figure}[t]
\centering \includegraphics[width = 86mm]{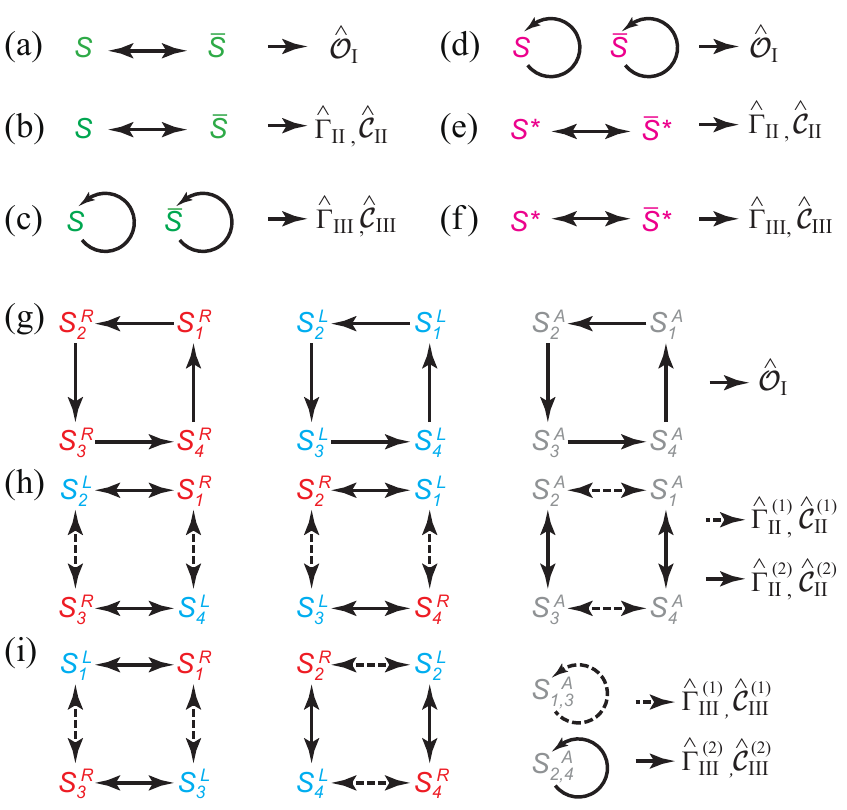}
\caption{ \label{fig:solitonsclass} 
[(a)--(c)] Transformations of SSH solitons under (a) class~I, (b) II, and (c) III operators.
[(d)--(f)] Transformations of RM solitons under (d) class~I, (e) II, and (f) III operators.
[(g)--(i)] Transformations of RC, LC, and AC solitons under (g) class~I, (h) II, and (i) III operators in the DC model.
See Table~\ref{table:transformation_soliton} for more information.
}
\end{figure}

We now investigate the roles of the class~I, II, and III operators 
among topological solitons using the low-energy effective continuum theory~\cite{brazovskii1980,RM1982, jackiw1983}. 
The results are summarized in Fig.~\ref{fig:solitonsclass} and Table~\ref{table:transformation_soliton}. See detail proofs in Appendix~\ref{appendix:Topological solitons}.

First, class~I operators endow cyclic equivalent relations to solitons
leading to a cyclic group.
For the SSH model, $\hat{G}$ shifts a soliton by a half unit cell in real space,
which transforms $S$ into $\bar S$ and vice versa [Fig.~\ref{fig:solitonsclass}(a)].
As $\hat{G}$ does not change soliton's physical properties,
it endows the equivalence relation between $S$ and $\bar S$:
Symbolically, $\bar{S} = \hat{G} S $ and $S = \hat{G} \bar{S} $.
As $S = \hat{G}^2 S$ and $\bar S = \hat{G}^2 \bar S $, SSH solitons
not only respect the $\mathbb{Z}_2$ transformation properties of the dimerized ground states under  $\hat{G}$
but also form a representation of a $\mathbb{Z}_2$ group, of which the generator is $\hat{G}$.
For the RM model, $\hat{G}^{\text{PT}}$ inverts a soliton and translates the inverted soliton by a half unit cell,
which transforms a soliton into itself having the same energy eigenvalue:
$S = \hat{G}^{\text{PT}} S $ and $\bar S = \hat{G}^{\text{PT}}\bar{S}$ [Fig.~\ref{fig:solitonsclass}(d)].
Thus, $\hat{G}^{\text{PT}}$ endows equivalent relations to RM solitons, as well.

Similarly, for the DC model, the class~I operator ($\hat{G}_y$) gives equivalence relations 
among solitons of the same chirality.
Under $\hat{G}_y$,
the four RC solitons are cyclically transformed into other RC solitons
having the same energy spectrum [Fig.~\ref{fig:solitonsclass}(g)].
Similarly, the four LC solitons and four AC solitons are cyclically transformed under $\hat{G}_y$ [Fig.~\ref{fig:solitonsclass}(g)].
Like the single-chain model,
each set of chiral solitons of the same chirality
not only respects the $\mathbb{Z}_4$ transformation properties of the dimerized ground states
but also forms the representation of a $\mathbb{Z}_4$ group, 
which can be useful in the topological operation~\cite{kim2017}.

Next, class~II operators endow the particle-hole dualities to solitons.
In the single-chain models, $\hat{\Gamma}_{\text{II}}$ effectively exchanges the sublattice potentials in every unit cell and translates a soliton by a half unit cell.
The soliton then transforms into the antisoliton with the opposite energy spectrum, and vice versa [Figs.~\ref{fig:solitonsclass}(b) and \ref{fig:solitonsclass}(e)].
Furthermore, the wave functions of the soliton and antisoliton ($\Psi_{S}$ and $\Psi_{\bar S}$)
are related by $\hat{\mathcal{C}}_{\text{II}}$
as $\Psi_{\bar S} \propto \hat{\mathcal{C}}_{\text{II}} \Psi_S=  i \sigma_x \Psi_S^*$, which endows the particle-hole duality in the quantum level.

Similarly, for the DC model, 
the class~II operator ($\hat{\Gamma}^{(i)}_{\text{II}}$, $\hat{\mathcal {C}}^{(i)}_{\text{II}}$)
transforms an RC soliton into an LC soliton having the opposite soliton energy states, and vice versa,
endowing the particle-hole duality between RC and LC solitons 
[Fig.~\ref{fig:solitonsclass}(h)].
For example,
$S_1^R= \hat{\Gamma}^{(2)}_{\text{II}} S_2^L$ [Fig.~\ref{fig:solitonsclass}(h)],
which explains the opposite spectra of the RC and LC solitons [Figs.~\ref{fig:soliton_spectra}(c) and \ref{fig:soliton_spectra}(d)].
The complete relations are shown in Fig.~\ref{fig:solitonsclass}(h).
It is noteworthy that due to the particle-hole duality,
solitons and antisolitons can be created or annihilated pairwise
like an ordinary particle-antiparticle pair.

Finally, the class~III operator can endow either self- or particle-hole dualities depending on the type of solitons.
Under the class~III operator ($\hat{\Gamma}_{\text{III}}$ or $\hat{\mathcal{C}}_{\text{III}}$), an SSH soliton is transformed into itself with the opposite energy level [Fig.~\ref{fig:solitonsclass}(c)].
As a result, the SSH soliton should have a symmetric spectrum, which allows a zero energy state only [Fig.~\ref{fig:soliton_spectra}(a)].
However, for the RM solitons, there is no allowed self-duality due to the sublattice symmetry breaking.
Rather, the class~III operator 
transforms $S^*$ to $\bar S^*$ with the opposite energy level, and vice versa,
endowing the particle-hole duality to the RM solitons [Figs.~\ref{fig:soliton_spectra}(b) and \ref{fig:solitonsclass}(f)].

For the DC model, the class~III operator endows the self-duality to
AC solitons as it does in the SSH model and particle-hole duality to a pair of RC and LC solitons as it does in the RM model
[Fig.~\ref{fig:solitonsclass}(i)].
For example,  $\hat{ \Gamma }^{(i)}_{\text{III}}$
transforms an AC soliton (RC soliton) into itself (LC soliton) with the opposite soliton energy states.
For AC solitons, this self-duality explains the symmetrically located midgap states [Fig.~\ref{fig:soliton_spectra}(e)].

\begin{table}
\centering 
\caption{
\label{table:topological_charge}
Relations among topological charges.
$Q_S$ and $Q_{\bar{S}}$ ($Q_{S^*}$ and $Q_{\bar{S}^*}$) are 
the topological charges of the soliton and antisoliton for the SSH (RM) model, respectively.
For the DC model, 
$Q_{S^k_{i}}$ ($k=R,L,A$) indicate the topological charges of RC, LC, and AC solitons.
Here $i, j =1,2,3,4$ and $S^k_{0,1}=S^k_{4,5}$.
}
{\renewcommand{\arraystretch}{0.5}%
\begin{tabular}{@{\extracolsep{15pt}} l c c c}
\hline
\hline
\\
Model & Class & Charge relation & Role \\
\\
\hline
\\
& I & $Q_S = Q_{\bar{S}}$ & Equivalence
\\
SSH & II & $Q_S = - Q_{\bar{S}}$ & Particle-hole
\\
& III & $Q_{S} = -Q_{S}$, $Q_{\bar{S}} = -Q_{\bar{S}}$ & Self-duality
\\
\\
\hline
\\
& I & $Q_{S^{*}} = Q_{S^{*}}$, $Q_{\bar{S}^{*}} = Q_{\bar{S}^{*}}$ & Equivalence
\\
RM & II & $Q_{S^{*}} = - Q_{\bar{S}^{*}}$ & Particle-hole
\\
& III & $Q_{S^{*}} = - Q_{\bar{S}^{*}}$ & Particle-hole
\\
\\
\hline
\\
& I & $Q_{S^k_i} = Q_{S^k_{j}}$  & Equivalence
\\
&II & $Q_{S^R_i} = - Q_{S^L_{i\pm 1}}$ & Particle-hole
\\
DC &II& $Q_{S^A_{i}} = - Q_{S^A_{i+1}} $ & Particle-hole
\\
& III & $Q_{S^R_i} = - Q_{S^L_{j}}$ & Particle-hole
\\
& III & $Q_{S^A_{i}} = - Q_{S^A_{i}} $ & Self-duality
\\
\\
\hline \hline
\end{tabular}}

\end{table}

\section{Topological properties of solitons}
\subsection{Topological charges}
\begin{figure*}[ht]
\centering \includegraphics[width=\linewidth]{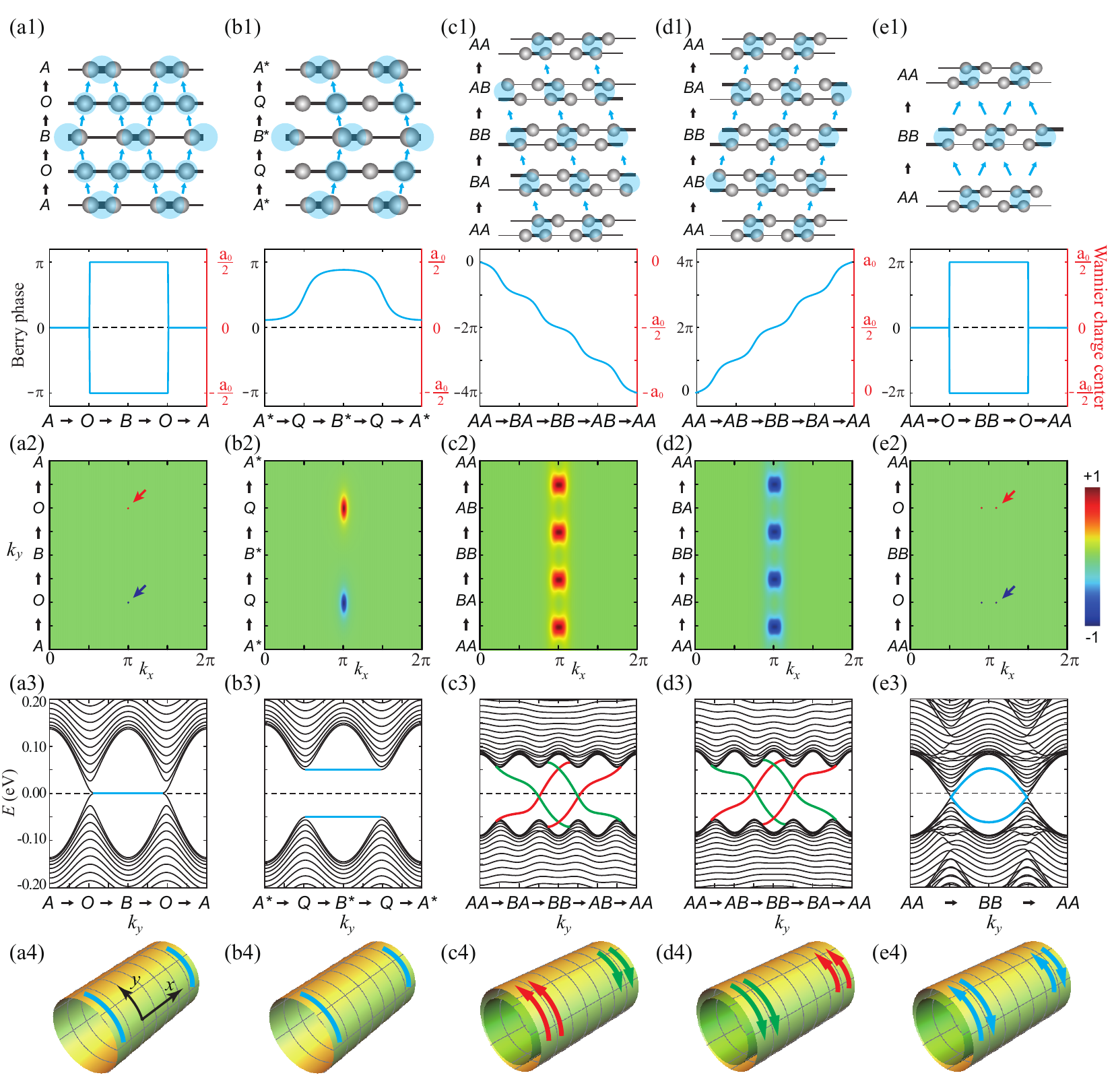} 
\caption{\label{fig:adiabatic_evolution} 
Adiabatic evolutions, topological charge pumpings, and chiral edge states.
[(a1)--(e1)]
Adiabatic evolutions (top panel) and  topological charge pumping (bottom panel) for the (a1) SSH, (b1) RM, (c1) RC, (d1) LC, and (e1) AC solitons.
Each adiabatic evolution is generated by transporting successive solitons along the adiabatic paths in the order parameter space [Figs.~\ref{fig:solitons}(f)--\ref{fig:solitons}(i)].
The movements of Wannier charge centers (cyan circles) are represented by the cyan arrows.
Each bottom graph shows the calculated Berry phase (left axis) and Wannier charge center (right axis) under one cycle of the adiabatic evolution.
In (a1), (b1), and (e1), there is no Thouless topological charge pumping.
In (c1) [(d1)], the quantized charges are pumped to the left (right).
[(a2)--(e2)]
Calculated 2D Berry curvature maps of the corresponding 2D effective systems,
where each adiabatic evolution is represented by $k_y$ momentum in the extra dimension. 
The color scale is normalized.
In (a2) and (e2), red and blue arrows indicate the singular points.
[(a3)--(e3)]
Calculated band spectra with open (closed) boundary condition along the $x$ direction ($y$ direction)
over the cylindrical geometries in (a4)--(e4).
Dispersions of edge states are indicated by blue, red, and green colors.
[(a4)--(e4)]
Schematics of  cylindrical geometries and edge states in real space.
On each cylinder, red and green arrows (blue lines and arrows) indicate chiral (trivial) edge states.
%
The chiral edge states for RC and LC solitons move oppositely in (c4) and (d4),
leading to the particle-hole duality between them.
In (e3) and (e4), the AC solitons have both right- and left-moving edge states at each side,
leading to the self-duality like the SSH solitons in (a3) and (a4).
} 
\end{figure*}

We now investigate the roles of the class~I, II, and III operators in the topological properties of topological solitons
using an adiabatic evolution and the corresponding effective two-dimensional (2D) systems.
An adiabatic evolution can be generated by transporting solitons very slowly along the adiabatic path [Figs.~\ref{fig:adiabatic_evolution}(a1)--\ref{fig:adiabatic_evolution}(e1)]
and the corresponding effective 2D Hamiltonian is obtained by taking the second dimension in the momentum space as the cyclic evolution~\cite{qi2008,cheon2015} [Figs.~\ref{fig:adiabatic_evolution}(a2)--\ref{fig:adiabatic_evolution}(e2)].
The corresponding topological charges of solitons can be calculated
through the generalized Goldstone-Wilczek formula or the partial phase-space Chern number~\cite{goldstone1981, qi2008}.
As a result, the class~I operators give the equivalent relations among topological soliton charges while the class~II operators do the particle-hole relations. The class~III operators can endow either particle-hole or self-duality depending on the model system.
The results are summarized in Table~\ref{table:topological_charge}.
See detail proofs in Appendix~\ref{appendix:topological charge}.

In the SSH model,
the soliton charge $Q_{S}$ and the antisoliton charge $Q_{\bar S}$
are obtained from the  adiabatic processes $A \rightarrow B$ and $B \rightarrow A$, respectively [Figs.~\ref{fig:adiabatic_evolution}(a1) and \ref{fig:adiabatic_evolution}(a2)].
Then class~I ($\hat{G}$) and II operators ($\hat{\Gamma}_{\text{II}}, \hat{\mathcal{C}}_{\text{II}}$) endow the equivalence relation ($Q_{S} = Q_{\bar S}$)
and the particle-hole relation ($Q_{S} = - Q_{\bar S}$), respectively.
On the other hand, class~III operator ($\hat \Gamma_{\text{III}}, \hat{\mathcal{C}}_{\text{III}}$) endows the self-duality ($Q_{X} = - Q_{X}$), where $X=S, \bar{S}$.
Combining these, because the soliton charge is defined up to modulo $\abs{e}$,
the SSH soliton charge is consistently  given by $ \pm \frac{1}{2} \abs{e}$ per spin~\cite{SSH1988}.

Similarly, in the RM model, 
the soliton charge $Q_{S^*}$ and the antisoliton charge $Q_{\bar S^*}$
can be obtained from the adiabatic processes $A^* \rightarrow B^*$ and $B^* \rightarrow A^*$, respectively [see Figs.~\ref{fig:adiabatic_evolution}(b1) and \ref{fig:adiabatic_evolution}(b2)].
First, the class~I operator ($\hat{G}^{\text{PT}}$) gives the equivalence relation: $Q_{S} = Q_{S}$ and $Q_{\bar S} = Q_{\bar S}$.
Second, like the SSH model, the class~II operator ($\hat{\Gamma}_{\text{II}}, \hat{\mathcal{C}}_{\text{II}}$) gives
the particle-hole relation of  $Q_{S^*} = - Q_{\bar{S}^*}$,
which is consistent with the known topological charge of the solitons in the RM model~\cite{RM1982}:
$Q_{S^*} = \frac{\abs{e}}{2} (1 - f)$ and $Q_{\bar{S}^*} = \frac{\abs{e}}{2} (1 +f) = - Q_{S^*} $ (mod $\abs{e}$),
where $f$ is the fractional charge deviated from the SSH soliton due to the sublattice symmetry breaking.

In the DC model,
the class~I operator ($\hat{G}_y$) imposes the equivalent relations such that 
the topological charges of chiral solitons of the same chirality are exactly the same.
In contrast, the class~II operator ($\hat{\Gamma}_{\text{II}}^{(i)}$, $\hat{\mathcal{C}}_{\text{II}}^{(i)}$) gives the particle-hole relation
to the topological charges of the RC and LC solitons such that $Q_{S^R_i} = - Q_{S^L_{i\pm 1}}$.
The class~II operator also endows the particle-hole relation to the topological charges of the AC solitons: $Q_{S^A_i} = - Q_{S^A_{i+1}}$.
The class~III operator can endow the self-duality to
the topological charges among AC solitons as it does in the SSH model ($Q_{S^A_i} = - Q_{S^A_{i}}$) and give particle-hole relation to a pair of RC and LC solitons as it does in the RM model ($Q_{S^R_i} = - Q_{S^L_{j}}$).
Combining these relations and the calculated Chern number~\cite{cheon2015} of $\pm 2$,
the topological charges of the RC and LC solitons are obtained:
$ Q_{S^R_i} = - Q_{S^L_i} = -\frac{1}{2}\abs{e}$ per spin.
Similarly, for the AC soliton, one can derive $Q_{S^A_i} = \pm Q_{S^A_i}$.  
Here $i, j =1,2,3,4$ and $S^k_{0,1}=S^k_{4,5}$.
As the soliton charge is defined up to modulo $\abs{2e}$ for the DC model,
the possible AC soliton charges are $Q^A_i = 0, \pm\abs{e}$ per spin.

\subsection{Topological charge pumping}
Using class~I, II, and III operators, we have found the equivalent and particle-hole relations among topological solitons.
Based on this finding, we now systematically discuss topological properties of the SSH, RM, and DC models.

Due to the equivalence and particle-hole duality, 
there is no topologically protected charge pumping in the SSH and RM models.
In the SSH model,
along the adiabatic evolution $A \rightarrow B \rightarrow A $, the Wannier charge centers~\cite{marzari1997}
split into two parts and then return to their original positions after one adiabatic cycle [Fig.~\ref{fig:adiabatic_evolution}(a1)].
Similarly, in the RM model, the Wannier charge center moves to the right and finally returns to its original position  [Fig.~\ref{fig:adiabatic_evolution}(b1)]
along the adiabatic evolution $A^* \rightarrow B^* \rightarrow A^* $.

On the other hand, in the DC model, the topological charge pumping is allowed during the adiabatic processes that correspond to four successive RC and LC solitons.
Along the adiabatic evolution $ AA \rightarrow BA \rightarrow BB \rightarrow AB \rightarrow AA $,
two electrons are pumped to the left
[Fig.~\ref{fig:adiabatic_evolution}(c1)]
while two are pumped to the right along the opposite adiabatic evaluation
$ AA \rightarrow AB \rightarrow BB \rightarrow BA \rightarrow AA $ [Fig.~\ref{fig:adiabatic_evolution}(d1)].
For the adiabatic process using AC solitons, however, there is no topological charge pumping like the SSH model [Fig.~\ref{fig:adiabatic_evolution}(e1)].

These features are clearly encoded in the phase-space Berry curvatures.
Due to the particle-hole duality,
the signs of the Berry curvature for two adiabatic processes ($A \rightarrow B$ and $B \rightarrow A$ that correspond to soliton and antisoliton, respectively) are opposite in the SSH model [Fig.~\ref{fig:adiabatic_evolution}(a2)], which leads to a zero total Chern number.
Similarly, the Berry curvatures for two adiabatic processes ($A^* \rightarrow B^*$ and $B^* \rightarrow A^*$ that correspond to soliton and antisoliton, respectively)  show the opposite signs in the RM model leading to a zero total Chern number as well.
Note that we take the limit of $m_z \rightarrow 0$ in the RM model to calculate Berry curvature in the gapless SSH model.
For the DC model, due to the particle-hole duality,
the Berry curvatures for the RC and LC solitons also lead to opposite total Chern numbers [Figs.~\ref{fig:adiabatic_evolution}(c2) and \ref{fig:adiabatic_evolution}(d2)].
For the AC soliton, 
the signs of the Berry curvature for the adiabatic processes ($AA \rightarrow BB$ and $BB \rightarrow AA$) are opposite like the SSH model, which leads to a zero total Chern number [Fig.~\ref{fig:adiabatic_evolution}(e2)].

\subsection{Soliton chirality}
Furthermore, the collaboration of the equivalence relation and particle-hole duality determines the existence of soliton chirality.
As soliton chirality is inherited from the chiral edge states of the quantum Hall insulator~\cite{haldane1988,cheon2015},
a sufficient condition is either the time-reversal symmetry breaking or a nonzero total Chern number.

However, for the SSH and RM models, 
adiabatic processes, which are generated by the soliton and antisoliton pair [Figs.~\ref{fig:adiabatic_evolution}(a1) and \ref{fig:adiabatic_evolution}(b1)], 
have a zero total Chern number
due to the particle-hole duality [Figs.~\ref{fig:adiabatic_evolution}(a2) and \ref{fig:adiabatic_evolution}(b2)] permitting no chirality to solitons.
The corresponding adiabatic evolution and the effective 2D Hamiltonians (Appendix~\ref{appendix:2D effective Hamiltonian}) have the time-reversal symmetry [Figs.~\ref{fig:adiabatic_evolution}(a1) and \ref{fig:adiabatic_evolution}(b1)], leading to no chiral edge states in the 2D cylindrical geometry [Figs.~\ref{fig:adiabatic_evolution}(a3) and \ref{fig:adiabatic_evolution}(b3)].

By contrast, two possible adiabatic processes,
which are generated by either RC solitons or LC solitons via the equivalent relations [Figs.~\ref{fig:adiabatic_evolution}(c1) and \ref{fig:adiabatic_evolution}(d1)],
have opposite total Chern numbers~\cite{cheon2015} of $\pm 2$
due to the particle-hole duality [Figs.~\ref{fig:adiabatic_evolution}(c2) and \ref{fig:adiabatic_evolution}(d2)] permitting the opposite chiralities to the solitons.
The corresponding effective 2D Hamiltonians  become Chern insulators with time-reversal symmetry breaking (Appendix~\ref{appendix:2D effective Hamiltonian}), which leads to the opposite chiral edge states in the 2D cylindrical geometry [Figs.~\ref{fig:adiabatic_evolution}(c3), \ref{fig:adiabatic_evolution}(c4), \ref{fig:adiabatic_evolution}(d3), and \ref{fig:adiabatic_evolution}(d4)].
For AC solitons, the recovered time-reversal symmetry 
makes the total Chern number zero [Fig.~\ref{fig:adiabatic_evolution}(e2)] leading to no chirality [Figs.~\ref{fig:adiabatic_evolution}(e3) and \ref{fig:adiabatic_evolution}(e4)] like SSH solitons [Figs.~\ref{fig:adiabatic_evolution}(a3) and \ref{fig:adiabatic_evolution}(a4)].

\subsection{Electrical charges of soliton and topological algebra}

Finally, we recover the spin degree of freedom
and discuss the electric charges and possible topological algebra of chiral solitons for potential topological information devices.
First, consider RC and LC solitons.
Because RC and LC solitons have a topological charge of $Q_i^R = - Q_i^L = - \abs{e}/2$ per spin,
RC and LC solitons have electric charges in integer multiples of $\abs{e}$ when considering the spin degree of freedom.
When the Fermi level lies at $E=0$,
an RC soliton (LC soliton) is negatively (positively) charged by $\abs{e}$ with no spin, acting like a spinless electron (hole): $q_{S^R} =  - \abs{e}$ and $q_{S^L} =  + \abs{e}$.
Here $q_{S^k}$ denotes the electric charge of solitons ($k=R,L, A$).

Next consider the AC solitons.
An AC soliton can have three electrical charges $q_{S^A} = 0$, $\pm 2 \abs{e}$,
because an AC soliton can have three topological charge values $Q^A_i = 0$, $\pm \abs{e}$ per spin
based on the symmetry analysis.
In the absence of interchain coupling, an AC soliton is composed of two SSH solitons (one for each chain).
When the charge-neutrality of the system is maintained,
an SSH soliton has only one electron state, which leads a spinful charge-neutral soliton.
However, in the presence of the interchain coupling, two soliton states are located above and below $E=0$, respectively [Fig.~\ref{fig:soliton_spectra}(e)].
Then two electrons occupy the lower energy level and an AC soliton becomes both chargeless and spinless.
In this case, in contrast to the SSH solitons, an AC soliton is a new extended state having no charge and no spin: $q_{S^A} =  0$.
Thus, this neutral AC soliton can survive very long time even when it interacts with other external defects having either charge or spin.
On the other hand, when $E_{\text{F}}$ is located between the upper (lower) soliton states and valence (conduction) band, 
the AC soliton has the electrical charge of $q_{S^A} =-2 \abs{e}$ ($+2 \abs{e}$) with no spin.

Furthermore, in the sense of topological operations,
the topological electric charges of chiral solitons respect $\mathbb{Z}_4$ topological algebra.
That is, the topological solitons can be added or subtracted in between and their topological charges satisfy the $\mathbb{Z}_4$ algebra during the corresponding addition or subtraction.
For example, an AC soliton ($S^A_1$) is equivalent to the addition of two successive RC solitons ($S^R_1$ and $S^R_2$) or LC solitons ($S^L_2$ and $S^L_3$)
as shown in the order parameter space [Figs.~\ref{fig:solitons}(g)--\ref{fig:solitons}(i)].
Then the electric charges automatically satisfy the algebra of  $q_{S^A} = 2 q_{S^R} =  2 q_{S^L}$ (mod $4 \abs{e}$), because $q_{S^A} =  \pm 2\abs{e}$, $q_{S^R} =  - \abs{e}$, and $q_{S^L} =  + \abs{e}$.

\section{Conclusion}
We have demonstrated a general framework that explains topological features of ground states as well as topological solitons in prototypical quasi-1D systems such as the SSH, RM, and DC models using the generalized time-reversal, particle-hole and chiral symmetry operators.
Combining time-reversal, particle-hole, chiral, and nonsymmorphic symmetry operators, we established three essential operators which are symmetry operators before the spontaneous symmetry breaking.
After spontaneous symmetry breaking,
the class~I and II operators connect degenerate ground states while the class~III operators give particle-hole symmetry to each ground state,
which is an extended Goldstone theorem to the 1D lattice systems.
For topological solitons, the class~I operators endow equivalent relations and cyclic group structures while the class~II and III operators do particle-hole relations.
Using these operators, we systematically described three distinct types of topological charge pumping and soliton chirality in the SSH, RM, and DC models.
Our work can be easily applied in various condensed matter systems as well as photonic crystal and cold atomic systems.

\appendix

\begin{table}[t]
\caption{
\label{appendix:table:basic_operators_single_chain_model}
Basic nonspatial and spatial operators for the SSH and RM models
in the Bloch basis and low-energy continuum theory.
$ \hat { G } \equiv \{E|\frac{a_0}{2}\} $ is the half-translation operator, $ \hat{\mathcal{P}} $ is the inversion operator with respect to the bonding center (Fig.~\ref{fig:models}), and $\hat K$ is the complex-conjugation operator.
}
{\renewcommand{\arraystretch}{0.6}%
\begin{tabular}{@{\extracolsep{8pt}} l c c c}
\hline
\hline
\\
Type & Operator  & Bloch  & Continuum  \\
\\
\hline
\\
&
$\hat{\mathcal{T}} $
&   $ \hat{K} \otimes ( k_x \rightarrow - k_x ) $ & $ \sigma_z  \hat{K}$ 
\\
Nonspatial &
$ \hat{ \mathcal{C} } $
&  $ \sigma_z \hat{K} \otimes ( k_x \rightarrow -k_x )$ 
& $ \hat{K} $ 
\\
&
$\hat{\Gamma} $ & $ \sigma_z $
& $ \sigma_z   $
\\
\\
\hline
\\
Spatial &
$\hat{G}  $
& $ e^{-i \frac{k_x a_0}{2}} \sigma_x $ & $  e^{ i \frac{\pi}{2}} \sigma_x$
\\
&
$\hat{\mathcal{P}}$
& $ \sigma_x \otimes (k_x \rightarrow -k_x )$ 
& $ \sigma_y \otimes (x \rightarrow -x)$
\\
\\
\hline \hline
\end{tabular}}

\end{table}

\begin{table}[t]
\caption{
\label{appendix:table:basic_operators_double_chain_model}
Basic nonspatial and spatial operators for the DC model
in the Bloch basis and low-energy continuum theory.
$\hat{\mathcal{T}}_{\text{D}}$, $\hat{ \mathcal{C}}_{\text{D}}$, and $\hat{\Gamma}_{\text{D}}$ are
the extended time-reversal, particle-hole, and chiral operators, respectively.
$\hat{G}^{(i)} \equiv \{ E | (-1)^{i+1} \frac{a_0}{2} \}^{(i)} $ is the fractional translation operator along the $i$th chain only.
$\hat{G}_y \equiv \{ M_y | - \frac{a_0}{4} \}$ is the glide reflection operator,
where $M_y$ is a mirror operator with respect to the $xz$ plane.
}
{\renewcommand{\arraystretch}{0.6}%
\begin{tabular}{@{\extracolsep{1pt}} l c c c}
\hline
\hline
\\
Type & Operator  & Bloch & Continuum  \\
\\
\hline
\\
&
$\hat{\mathcal{T}}_{\text{D}} $
&   $\begin{pmatrix} \mathbf{1} & 0  \\ 0 & \mathbf{1} \end{pmatrix} \hat  K \otimes (k_x \rightarrow -k_x) $ 
& $\begin{pmatrix} \sigma_z & 0  \\ 0 & e^{ - i \frac{\pi}{2}} \sigma_z  \end{pmatrix} \hat  K $ 
\\
\\
Non &
$ \hat{ \mathcal{C}}_{\text{D}} $
& $\begin{pmatrix}  \sigma_z & 0  \\ 0 &  \sigma_z  \end{pmatrix} \hat  K \otimes (k_x \rightarrow -k_x) $ 
& $\begin{pmatrix} \mathbf{1} & 0  \\ 0 &e^{ - i \frac{\pi}{2}}  \mathbf{1}  \end{pmatrix} \hat  K $ 
\\
\\
&
$\hat{\Gamma}_{\text{D}} $ & $\begin{pmatrix} \sigma_z & 0 \\ 0 & \sigma_z \end{pmatrix}$
& $\begin{pmatrix} \sigma_z & 0 \\ 0 & \sigma_z \end{pmatrix}$
\\
\\
\hline
\\
&
$\hat{G}^{(1)}  $
&   $\begin{pmatrix} e^{-  i \frac{k_x a_0}{2}} \sigma_x & 0  \\ 0 & \mathbf{1}  \end{pmatrix}  $
&   $\begin{pmatrix}  e^{ i \frac{\pi}{2}}  \sigma_x & 0  \\ 0 & \mathbf{1}  \end{pmatrix}  $
\\
\\
Spatial &
$\hat{G}^{(2)} $
&   $\begin{pmatrix} \mathbf{1} &  0  \\  0 & e^{  i \frac{k_x a_0}{2}} \sigma_x  \end{pmatrix}  $
&  $\begin{pmatrix} \mathbf{1} &  0  \\  0 & e^{- i \frac{\pi}{2}}  \sigma_x  \end{pmatrix}  $
\\
\\
&
$\hat{G}_y $
& $ e^{ i \frac{k_x a_0}{4}} \begin{pmatrix} 0 & \sigma_x  \\ \mathbf{1} & 0  \end{pmatrix}  $
& $ e^{- i \frac{\pi}{4}} \begin{pmatrix} 0 & \sigma_x  \\ \mathbf{1} & 0  \end{pmatrix}  $
\\
\\
\hline \hline
\end{tabular}}

\end{table}

\section{Class~I, II, and III operators}
\label{appendix:classI_II_IIIoperators}
In Appendix \ref{appendix:classI_II_IIIoperators}, we explicitly present the class~I, II, and III operators.
The class~I, II, and III operators are constructed using the nonspatial and spatial operators which are
listed in Table~\ref{appendix:table:basic_operators_single_chain_model} and \ref{appendix:table:basic_operators_double_chain_model}.
The explicit representations of the class~I, II, and III operators are listed in Table~\ref{appendix:table:operators_single_chain_model} and \ref{appendix:table:operators_double_chain_model}.
Using these explicit forms, one can easily prove the symmetry transformation properties of both Hamiltonians and their ground states under the class~I, II, and III operators shown in
Table~\ref{table:transformation_hamiltonian}.

\begin{table}[h]
\caption{
\label{appendix:table:operators_single_chain_model}
Class~I, II, and III operators for the SSH and RM models in the Bloch basis and low-energy continuum theory.
$\hat{G} \equiv \{E|\frac{a_0}{2}\} $ is the half-translation operator,
$\hat{G}^{ \text{PT}} \equiv  \hat{G} \hat{\mathcal{P}}\hat{\mathcal{T}}$ is the PT-symmetric half-translation operator,
$\hat{\Gamma}^{\text{PT}}  \equiv \hat{\Gamma} \hat{\mathcal{P}} \hat{ \mathcal{ T } } $ is the PT-symmetric chiral operator, and
$ \hat{\mathcal{C}}^{\text{PT}}  \equiv \hat{\mathcal {C}} \hat{\mathcal{P}} \hat{ \mathcal{ T } }  $ is the PT-symmetric charge-conjugation operator.
$\hat{\mathcal{I}}_{k} \equiv ( k_x \rightarrow -k_x )$ and $\hat{\mathcal{I}}_{x} \equiv ( x \rightarrow -x )$.
}
{\renewcommand{\arraystretch}{0.6}%
\begin{tabular}{@{\extracolsep{6pt}} l c c c c}
\hline
\hline
\\
Model & Class & Operator  & Bloch & Continuum  \\
\\
\hline
\\
& $\hat{\mathcal{O}}_\text{I} $ & $ \hat{G}$
& $ e^{-i \frac{k_x a_0}{2}} \sigma_x $ & $ e^{ i \frac{\pi}{2}} \sigma_x$
\\
& $\hat{\Gamma}_\text{II}$ & $  \hat{G} \hat{\Gamma} $
& $- i e^{-i\frac{k_x a_0}{2}}  \sigma_y$ & $  \sigma_y$ 
\\
SSH
& $\hat{\mathcal{C}}_\text{II}$ & $ \hat{G} \hat{\mathcal{C}} $
& $ - i e^{-i\frac{k_x a_0}{2}} \sigma_y \hat{K} \otimes \hat{\mathcal{I}}_{k} $ 
& $  e^{ i \frac{\pi}{2}} \sigma_x \hat{K}$ 
\\
& $\hat{\Gamma}_\text{III}$ & $\hat{\Gamma}$
& $\sigma_z $ & $\sigma_z $
\\
& $\hat{\mathcal{C}}_\text{III}$ &  $\hat{\mathcal{C}}$
& $ \sigma_z \hat{K} \otimes \hat{\mathcal{I}}_{k} $ & $ \hat{K}$
\\
\\
\hline
\\
& $\hat{\mathcal{O}}_\text{I}$ & $\hat{G}^{ \text{PT}} $
& $-e^{-i \frac{k_x a_0}{2}} \hat{K}$ 
& $ - \hat{K} \otimes \hat{\mathcal{I}}_{x} $ 
\\
& $\hat{\Gamma}_\text{II}$ & $\hat{G} \hat{\Gamma} $
& $- i e^{-i\frac{k_x a_0}{2}}  \sigma_y$ & $  \sigma_y$ 
\\
RM
& $\hat{\mathcal{C}}_\text{II}$ & $ \hat{G} \hat{\mathcal{C}} $
& $ - i e^{-i\frac{k_x a_0}{2}} \sigma_y \hat{K} \otimes \hat{\mathcal{I}}_{k} $ 
& $  e^{ i \frac{\pi}{2}} \sigma_x \hat{K}$ 
\\
& $\hat{\Gamma}_\text{III}$ & $\hat{\Gamma}^{\text{PT}}   $
& $ i\sigma_y \hat{K}$
& $ -\sigma_y \hat{K} \otimes \hat{\mathcal{I}}_{x} $
\\
& $\hat{\mathcal{C}}_\text{III}$ &  $ \hat{\mathcal{C}}^{\text{PT}}  $
& $ i \sigma_y \otimes \hat{\mathcal{I}}_{k} $ & $ -i \sigma_x  \otimes \hat{\mathcal{I}}_{x} $
\\
\\
\hline \hline
\end{tabular}}

\end{table}

\begin{table*}
\caption{
\label{appendix:table:operators_double_chain_model}
Class~I, II, and III operators for the DC model in the Bloch basis and low-energy continuum theory.
}
{\renewcommand{\arraystretch}{0.6}%
\begin{tabular}{@{\extracolsep{40pt}} l c c c c}
\hline
\hline
\\
Model & Class &Operator & Bloch & Continuum  \\
\\
\hline
\\
&  $\hat{\mathcal{O}}_\text{I} $ & $\hat{G}_y  $
& $ e^{ i \frac{k_x a_0}{4}} \begin{pmatrix} 0 & \sigma_x  \\ \mathbf{1} & 0  \end{pmatrix}  $
& $ e^{- i \frac{\pi}{4}} \begin{pmatrix} 0 & \sigma_x  \\ \mathbf{1} & 0  \end{pmatrix}  $
\\
\\
& $\hat{\Gamma}^{(1)}_{\text{II}}$ & $ \hat{G}^{(1)} \hat{\Gamma}_{\text{D}}$
& $\begin{pmatrix} -i e^{- i \frac{k_x a_0} {2} }\sigma_y & 0  \\ 0 &  \sigma_z  \end{pmatrix}$ 
& $ \begin{pmatrix} \sigma_y & 0  \\ 0 & \sigma_z  \end{pmatrix} $ 
\\
\\
& $\hat{\Gamma}^{(2)}_{\text{II}}$ & $ \hat{G}^{(2)} \hat{\Gamma}_{\text{D}} $
& $ \begin{pmatrix} \sigma_z & 0  \\ 0 & - i e^{+ i \frac{k_x a_0} {2} } \sigma_y  \end{pmatrix} $ 
& $ \begin{pmatrix} \sigma_z & 0  \\ 0 & -\sigma_y  \end{pmatrix} $ 
\\
\\
& $ \hat{\mathcal{C}}^{(1)}_{\text{II}} $ & $ \hat{G}^{(1)} \hat{ \mathcal{C}}_{\text{D}} $
& $ \begin{pmatrix} - i e^{ - i \frac{k_x a_0} {2} }\sigma_y & 0  \\ 0 & \sigma_z  \end{pmatrix} \hat  K \otimes (k_x \rightarrow -k_x) $ 
& $ i \begin{pmatrix} \sigma_x & 0  \\ 0 & -  \mathbf{1}  \end{pmatrix} \hat  K  $ 
\\
\\
DC
& $ \hat{\mathcal{C}}^{(2)}_{\text{II}} $ & $ \hat{G}^{(2)} \hat{ \mathcal{C}}_{\text{D}}  $
& $ \begin{pmatrix} \sigma_z & 0  \\ 0 & -i  e^{ + i \frac{k_x a_0} {2} } \sigma_y  \end{pmatrix} \hat  K \otimes (k_x \rightarrow -k_x) $ 
& $ \begin{pmatrix} \mathbf{1} & 0  \\ 0 & -\sigma_x \end{pmatrix} \hat  K $ 
\\
\\
& $\hat{\Gamma}^{(1)}_{\text{III}}$ & $\hat{\Gamma}_{\text{II}}^{(1)} \hat{G}_y $
& $ e^{ i \frac{k_x a_0}{4}} \begin{pmatrix} 0 & - e^{ - i \frac{k_x a_0}{2} }\sigma_z  \\ \sigma_z & 0  \end{pmatrix}$
& $ e^{- i \frac{\pi}{4}} \begin{pmatrix} 0 &  e^{ - i \frac{\pi}{2} }\sigma_z  \\ \sigma_z & 0  \end{pmatrix} $
\\
\\
& $\hat{\Gamma}^{(2)}_{\text{III}}$ &  $ \hat{G}_y^{-1} \hat{\Gamma}_{\text{II}}^{(2)}  $
& $ - i e^{ - i \frac{k_x a_0}{4}} \begin{pmatrix} 0 & \sigma_y  \\ e^{ i \frac{k_x a_0}{2} } \sigma_y & 0  \end{pmatrix} $ 
& $  e^{ i \frac{\pi}{4}} \begin{pmatrix} 0 & 
- \sigma_y  \\ -i \sigma_y & 0  \end{pmatrix} $
\\
\\
& $ \hat{\mathcal{C}}^{(1)}_{\text{III}} $ &  $ \hat{\mathcal{C}}_{\text{II}}^{(1)} \hat{G}_y $
& $ e^{ i \frac{k_x a_0}{4}} \begin{pmatrix} 0 & - e^{ - i \frac{k_x a_0}{2} }\sigma_z  \\ \sigma_z & 0  \end{pmatrix} 
\hat  K \otimes (k_x \rightarrow -k_x)$
& $ i e^{ i \frac{\pi}{4}}  \begin{pmatrix} 0 &  1  \\ - 1 & 0  \end{pmatrix} \hat{K} $
\\
\\
& $ \hat{\mathcal{C}}^{(2)}_{\text{III}} $ &  $ \hat{G}_y^{-1}  \hat{\mathcal{C}}_{\text{II}}^{(2)} $
& $- i e^{- i \frac{k_x a_0}{4}} \begin{pmatrix} 0 & \sigma_y  \\ e^{ i \frac{k_x a_0}{2} } \sigma_y & 0  \end{pmatrix}
\hat  K \otimes (k_x \rightarrow -k_x) $ 
& $  e^{ i \frac{\pi}{4}} \begin{pmatrix} 0 & 
- \sigma_x  \\  \sigma_x & 0  \end{pmatrix} \hat{K} $
\\
\\
\hline \hline
\end{tabular}}

\end{table*}

\section{Proofs of the transformation properties of topological solitons under class~I, II, and III operators}
\label{appendix:Topological solitons}

In Appendix \ref{appendix:Topological solitons}, we explicitly prove the transformation properties among topological solitons under the class~I, II, and III operators using the low-energy effective Hamiltonians.

\subsection{Single-chain model}
Consider a soliton solution having a wave function $\Psi(x)$,
a dimerization profile $\Delta(x)$, and an energy eigenvalue $E$.
Because the soliton solution can be obtained using the low-energy effective continuum Hamiltonian~\cite{brazovskii1980,takayama1980,RM1982, jackiw1983}, the solution satisfies the following eigenvalue equation:
\begin{equation}
\mathsf{H}_{\text{SSH}} (-i \partial_x, \Delta(x)) \Psi(x)  = E \Psi(x).
\end{equation}
On the other hand, 
the low-energy effective Hamiltonian $\mathsf{H} _{ \text{SSH} }$ transforms under an operator $\hat{O}$ as
\begin{equation}
\hat{O} \mathsf{H} _{ \text{SSH} } (-i \partial_x, \Delta(x))  \hat{O}^{-1} = \eta \mathsf{H}_{ \text{SSH} } (-i \partial_x, \Delta\rq{}(x))
\end{equation}
as shown in Table \ref{table:transformation_hamiltonian}.
Therefore, the transformed wave function $\Psi'(x) \equiv  \hat{O} \Psi(x) $ is also a solution that satisfies the following eigenvalue equation:
\begin{equation}
\mathsf{H}_{\text{SSH}} (-i \partial_x, \Delta'(x)) \Psi'(x)  = E' \Psi'(x) ,
\end{equation}
where  $\Delta'(x) $ and $E' = \eta E$ with $\eta = \pm 1$ are the dimerization profile and energy spectrum of the transformed soliton.
Depending on the transformed dimerization profile and energy spectrum, the transformed soliton can be determined either as a soliton or an antisoliton.

Let us see the role of the class~I, II, and III operators on the topological solitons in the SSH model.
First, under the class~I operator $\hat G$, the transformed soliton has $\Delta'(x) = - \Delta(x) $ and $E' = E$.
Since the sign of the dimerization displacement profile $\Delta(x)$ is reversed,
$\hat{G}$ transforms $S$ into $\bar{S}$ with the same energy eigenvalue, and vice versa;
symbolically, $\hat {G}  S = \bar{S} $ and $  \hat {G}  \bar{S} =  S$ with $E_{S} = E_{\bar S}$.
Therefore, the class~I operator endows an equivalent relation between solitons in the SSH model.
Moreover, if we repeatedly apply $\hat{G}$ two times to a soliton, it returns to itself:
$ \hat {G}^2  S = S $ and $ \hat {G}^2  \bar{S} = \bar S$.
Thus, the soliton and antisoliton form an irreducible representation for the $\mathbb{Z}_2$ group of which the generator is $\hat G$.

Next, under the class~II operator $\hat{\mathcal{C}}_{\text{II}}$ (or  $\hat{\Gamma}_{\text{II}}$), 
the transformed soliton has $\Delta'(x) = - \Delta(x) $ and $E' = - E$.
Then $\hat{\mathcal{C}}_{\text{II}}$  transforms $S$ into $\bar{S}$ with the reversed energy eigenvalue, and vice versa:
symbolically, $ \hat{\mathcal{C}}_{\text{II}}  S  =  \bar{S}  $ and $ \hat{\mathcal{C}}_{\text{II}}  \bar{S}  =  S $  with $E_{S} = - E_{\bar S}$.
In particular, the wave functions of the soliton and antisoliton form a particle-hole pair under $\hat{\mathcal{C}}_{\text{II}}$. 
Mathematically, 
$
\Psi_{\bar S}(x) \propto \hat{\mathcal{C}}_{\text{II}} \Psi_S(x) =  i \sigma_x \Psi_S^*(x).
$

Finally, under the class~III operator $\hat{\mathcal{C}}$ (or, equivalently,  $\hat{\Gamma}$),
the transformed soliton has   $\Delta'(x) = \Delta(x) $ and $E' = - E$.
Thus, a soliton transforms into itself with the opposite energy eigenvalue,
which allows the zero energy state:
symbolically, $ \hat{\mathcal{C}} S =   S $ and $  \hat{\mathcal{C}}  \bar S = \bar S $ with $E_{S} = - E_{S}=0$ and  $E_{\bar S} = - E_{\bar S}=0$.
In this sense, the class~III operator $\hat{\mathcal{C}}$ (or, equivalently, $\hat{\Gamma}$) endows the self-duality (or, equivalently, the particle-hole symmetric spectra) to each soliton and antisoliton.

Similar arguments can be applied to the transformation properties of topological solitons for the RM model.
Let a wave function $\Psi(x)$ be a soliton solution that has a dimerization profile $\Delta(x)$ and an energy eigenvalue $E$.
As shown in Table \ref{table:transformation_hamiltonian},
the low-energy effective Hamiltonian $\mathsf{H} _{ \text{RM} }$ transforms under an operator $\hat{O}$ as 
\begin{equation*}
\hat{O} \mathsf{H} _{ \text{RM} } (-i \partial_x, \Delta(x), m_z)  \hat{O}^{-1} = \eta \mathsf{H}_{ \text{RM} } (-i \partial_x,  \Delta\rq{}(x), m_z).
\end{equation*}
Thus, the transformed wave function $\Psi'(x) \equiv  \hat{O} \Psi(x) $ is also a solution that satisfies the following eigenvalue equation:
\begin{equation}
\mathsf{H}_{\text{RM}} (-i \partial_x, \Delta'(x), m_z ) \Psi'(x)  = E' \Psi'(x) ,
\end{equation}
where  $\Delta'(x) $ and $E' = \eta E$ are the dimerization profile and energy spectrum of the transformed soliton.
Depending on the dimerization profile, the transformed soliton can be either a soliton or an antisoliton.

First, under the class~I operator $ \hat{G}^{ \text{PT} }$,
the transformed soliton has  $\Delta'(x) = - \Delta(-x) = \Delta(x) $ and $E' = E$,
where the dimerization profile is explicitly inverted in real space due to the inversion operator $\hat{\mathcal{P}}$.
Since the dimerization displacement profile $\Delta(x)$ does not change,
$ \hat{G}^{ \text{PT} } $ transforms a soliton into itself having the same energy eigenvalue:
symbolically, $ \hat{G}^{ \text{PT} } S^* =    S^*$ and $ \hat{G}^{ \text{PT} } \bar S^* =   \bar S^*$.
In this sense, the class~I operator endows equivalent relations to solitons for the RM model.

Next, the class~II operator $\hat{\mathcal{C}}_{\text{II}}$ (or  $\hat{\Gamma}_{\text{II}}$) acts in the same way as it does for the SSH model,
endowing the particle-hole duality.
Symbolically, $  \hat{\mathcal{C}}_{\text{II}} {S}^* =  \bar S^*$ and $ \hat{\mathcal{C}}_{\text{II}}  \bar S^* = {S}^*$ with $E_{S^*} = - E_{\bar S^*}$.

Finally, under the class~III operator $\hat{\mathcal{C}}^{\text{PT}}$ (or, equivalently,  $\hat{\Gamma}^{\text{PT}}$), 
the transformed soliton has $\Delta'(x) = - \Delta(x) $ and $E' = - E$.
Thus, $\hat{\mathcal{C}}^{\text{PT}}$ transforms a soliton into an antisoliton with the opposite energy eigenvalue, and vice versa:
symbolically,  $ \hat{\mathcal{C}}^{\text{PT}} S^* =    \bar S^* $ and $ \hat{\mathcal{C}}^{\text{PT}} \bar S^* =   S^*$ with $E_{S^*} = - E_{\bar S^*}$.
Therefore, the class~III operator in the RM model gives the particle-hole duality unlike the class~III operator in the SSH model.
Note that the difference of the role of the class~III operators in the SSH and the RM model is attributed to the sublattice symmetry breaking;
the sublattice symmetry breaking in the RM model gives distinct energy spectra to the soliton and the antisoliton, which prohibits the self-duality.

\subsection{Double-chain model}

The solutions of the chiral solitons in the DC model can be obtained using the low-energy effective Hamiltonian.
Let us consider the wave function of a chiral soliton $ \Psi (x) $ that satisfies the following eigenvalue equation:
\begin{eqnarray}
\mathsf{H} ( \Delta^{(1)} (x) , \Delta^{(2)} (x) ) \Psi (x) = E \Psi (x),
\end{eqnarray}
where $\Delta^{(1)} (x)$ and $\Delta^{(2)} (x)$ are dimerization displacement profiles and $E$ is the energy eigenvalue for the chiral soliton.
As shown in Table \ref{table:transformation_hamiltonian},
the low-energy effective Hamiltonian $\mathsf{H} _{ \text{DC} }$ transforms under an operator $\hat{O}$ as
\begin{equation*}
\hat{O} \mathsf{H} _{ \text{DC} } (\Delta^{(1)}(x), \Delta^{(2)}(x) )  \hat{O}^{-1} = \eta \mathsf{H}_{ \text{DC} }(\Delta'^{(1)}(x), \Delta'^{(2)}(x) ).
\end{equation*}
Then the transformed wave function $\Psi'(x) \equiv  \hat{O} \Psi(x) $ is also a soliton solution that satisfies the following eigenvalue equation:
\begin{equation}
\mathsf{H}_{\text{DC}} (\Delta'^{(1)}(x), \Delta'^{(2)}(x) ) \Psi'(x)  = E' \Psi'(x) ,
\end{equation}
where  $(\Delta'^{(1)}(x), \Delta'^{(2)}(x) ) $ and $E' = \eta E $ are the dimerization profiles and energy spectrum of the transformed soliton.
Here $\eta= \pm1$.
Depending on the dimerization profile, the transformed chiral soliton can be RC, LC, or AC solitons.

First, under the class~I operator $ \hat{ G}_y $, the transformed soliton has $\Delta'^{(1)}(x) = - \Delta^{(2)}(x) $,  $\Delta'^{(2)}(x) = \Delta^{(1)}(x)$, and $ E' = E$.
Thus, $ \hat{ G}_y $ transforms a chiral soliton into another chiral soliton of the same chirality having the same energy eigenvalue.
Symbolically, we find that
\begin{eqnarray}
S_1^{k} &=& \hat{G}_y S_4^{k} =  \hat{G}_y^2 S_3^{k} = \hat{G}_y^3 S_2^{k} = \hat{G}_y^4 S_1^{k}
\label{seq:double_Gy_01}
\end{eqnarray}
for $k=R,L,A$.
Thus, $ \hat{ G}_y $ endows the equivalence relation to the chiral solitons of the same chirality
and hence we denote the energy eigenvalues as $E^{\text{RC}}$, $E^{\text{LC}}$, and $E^{\text{AC}}$ for RC, LC, and AC solitons, respectively.

Next, under the class~III operator $\hat{\mathcal{C}}^{(i)}_{\text{III}}$ (or, equivalently,  $\hat{\Gamma}^{(i)}_{\text{III}}$), 
the  transformed chiral soliton has the following dimerization profiles and energy eigenvalue:
\begin{eqnarray*}
\Delta'^{(1)} (x) & = & (-1)^{i+1} \Delta^{(2)} (x),\\
\Delta'^{(2)} (x) & = & (-1)^{i+1} \Delta^{(1)} (x),
\end{eqnarray*}
and $E'= - E$.
From this, one can find two types of transformations.
First, an AC soliton can transform into itself under $\hat{\mathcal{C}}^{(i)}_{\text{III}} $ [or, equivalently,  $\hat{\Gamma}^{(i)}_{\text{III}} $].
For example, consider the AC soliton $ S_1^A$  having $\Delta^{(1)} = \Delta^{(2)}$ and energy eigenvalue $E^{\text{AC}}$.
Under $\hat{\mathcal{C}}^{(1)}_{\text{III}} $, this AC soliton transforms into itself with the opposite eigenvalue $-E^{\text{AC}}$,
which implies the AC soliton has a particle-hole symmetric spectra $\pm E^{\text{AC}}$,
leading to a self-duality.
In this sense, the class~III operator $\hat{\mathcal{C}}^{(i)}_{\text{III}} $ endows the self-duality to each AC soliton.

On the other hand, under the class~III operator, an RC soliton can transform to an LC soliton with the opposite energy eigenvalues, and vice versa.
For example, the RC soliton $ S_1^R $ transforms into the LC soliton $ S_3^L $  under $ \hat{\mathcal{C}}^{(1)}_{\text{III}} $:
symbolically, $ S_3^L  = \hat{\mathcal{C}}^{(1)}_{\text{III}}   S_1^R $ with $E^{\text{RC}} = -E^{\text{LC}}$.
In this sense, the class~III operator $\hat{\mathcal{C}}^{(i)}_{\text{III}} $ endows the particle-hole duality.
Therefore, the class~III operator endows the self-duality to AC solitons as it does in the SSH model
and the particle-hole duality to RC and LC soliton pairs as it does in the RM model.

Finally, under the class~II operator $ \hat{\mathcal{C}}^{(i)}_{\text{II}} $ [or  $ \hat{\Gamma}^{(i)}_{\text{II}} $],
the transformed chiral soliton has $\Delta'^{(1)}(x) = (-1)^{i} \Delta^{(1)}(x)$, $\Delta'^{(2)}(x) = (-1)^{i+1} \Delta^{(2)}(x)$ and $ E' = - E$.
From this, one can find two types of transformations:
An RC soliton transforms into an LC soliton  with the opposite energy eigenvalues, and vice versa. On the other hand, 
an AC soliton transforms into another AC soliton with the same energy spectra because an AC soliton has the particle-hole symmetric energy spectra.
For example, under $ \hat{\mathcal{C}}^{(1)}_{\text{II}}$,
the RC soliton $ S_1^R $ transforms to the LC soliton $ S_2^L $
and the AC soliton $ S_2^A$ transforms to the AC soliton $ S_1^A $:
symbolically, $ \hat{\mathcal{C}}^{(1)}_{\text{II}} S_1^R  = S_2^L  $ with $E^{\text{RC}} = -E^{\text{LC}}$ and
$ \hat{\mathcal{C}}^{(1)}_{\text{II}} S_2^A  = S_1^A $.
For both cases, the original and transformed wave functions [$\Psi_{\text{O}}(x) $ and $\Psi_{\text{T}}(x)$]
satisfy the particle-hole relation under $ \hat{\mathcal{C}}^{(1)}_{\text{II}}$.
Therefore, the class~II operator endows the particle-hole duality between chiral solitons.

\section{Topological charges of solitons} \label{appendix:topological charge}
In this section, we explicitly prove the roles of the class~I, II, and III operators in the soliton charges for the SSH, RM, and DC models
using the corresponding adiabatic evolution and the corresponding partial Chern number~\cite{goldstone1981,cheon2015,qi2008}.

The corresponding adiabatic evolution can be generated by transporting solitons very slowly
and is represented by the time-dependent phase-space Hamiltonian.
For the single- and double-chain models, the phase-space Hamiltonians are given by
\begin{eqnarray}
\mathsf{H}_{\text{single}} [ k_x, t ] & =&  \mathsf{H}_{\text{single}} [ k_x, \Delta(t), m_z(t) ], \\
\mathsf{H}_{\text{DC}} [ k_x, t ] & = &  \mathsf{H}_{\text{DC}} [ k_x, \Delta^{(1)}(t), \Delta^{(2)}(t) ],
\end{eqnarray}
where $\Delta(t)$'s and $m_z(t)$ are time-dependent functions 
that satisfy the periodic boundary condition $\mathsf{H}[ k_x, t  + T] = \mathsf{H}[ k_x, t ] $ with a period $T$.
%

Basically, the topological charge of a topological soliton can be calculated
through the generalized Goldstone-Wilczek formula or the phase-space Chern number~\cite{goldstone1981,cheon2015,qi2008}.
That is, the partial phase-space Chern number $ C_{\text{partial}}$ under an adiabatic process
is equal to the topological charge $Q$ carried by the corresponding topological soliton
(or $ Q = - \abs{e} C_{\text{partial}}$).
The  partial phase-space Chern number from the initial time $t_i$ to the final time $t_f$ is defined as 
\begin{eqnarray} 
C_{\text{partial}} = \frac{i}{ 2 \pi} \sum_{n=\text{occ}} \int_{\text{BZ}} d k_x \int_{t_i}^{t_f}  dt ~\Omega_n,
\end{eqnarray}
where the summation is done over the occupied bands and 
$ \Omega_n =  \Braket{ \partial_{k_x} u_n |  \partial_{t} u_n } - \Braket{ \partial_{t} u_n  |  \partial_{k_x} u_n } $
is the phase-space Berry curvature.
From now on, we will omit $dk_x$ in the integral and use following notation for simplicity:
\begin{eqnarray}
\Omega_n (\hat{\mathcal{O}}, t) \equiv
&& \Braket{ \partial_{k_x} \hat{\mathcal{O}} u_n (k_x, t) |  \partial_{t} \hat{\mathcal{O}} u_n (k_x, t) } \\
&&
~~~~ - \Braket{ \partial_{t} \hat{\mathcal{O}} u_n (k_x, t)  |  \partial_{k_x} \hat{\mathcal{O}} u_n (k_x, t) }, \nonumber
\end{eqnarray}
where $\hat{\mathcal{O}}$ is an operator.
Note that the total Chern number $C_{\text{total}}$ is defined for one full cycle.

\subsection{Single-chain model}
Let us see consider the roles of the glide operator $\hat{G}$ (class~I)
and the  nonsymmorphic chiral operator $\hat{\Gamma}_{\text{II}}$ (class~II) 
on the topological charges of solitons in the SSH model.
Because the topological number does not depend on the details of the adiabatic process,
without loss of generality, we choose the adiabatic process along the straight lines in the order parameter space as shown in Figs.~\ref{fig:solitons}(f) and \ref{fig:adiabatic_evolution}(a1).
This adiabatic process respects the transformation properties which are imposed by $\hat{G}$ and $\hat{\Gamma}_{\text{II}}$.
Therefore, the phase-space Hamiltonian satisfies the following relations:
\begin{eqnarray} \label{seq:adiabatic_SSH_01}
\mathsf{H}_{\text{SSH}} \left[k_x, \frac{T}{2} +t \right]
 =    \hat{G}  \mathsf{H}_{\text{SSH}} [k_x, t ] \hat{G}^{-1} \label{eq:Adiabatic_Glide}
\end{eqnarray}
and
\begin{eqnarray}
&&
\mathsf{H}_{\text{SSH}} \left[k_x,  (2m-1) \frac{T}{4} + t\right]   \label{seq:adiabatic_SSH_02}
\\
&& ~~~~~~~~~~~~~~~~~~~~
= - \hat{\Gamma}_{\text{II}}    \mathsf{H}_{\text{SSH}} \left[ k_x,  (2m-1) \frac{T}{4} - t \right] (\hat{\Gamma}_{\text{II}}  )^{-1}, 
\nonumber
\end{eqnarray}
where $m \in \mathbb{Z},~~ t \in [0, \frac{T}{4} ] $.
Note that we choose such that $\mathsf{H}_{\text{SSH}}[ k_x, t = 0 ] $ and $\mathsf{H}_{\text{SSH}}[ k_x, t = T/2 ] $ correspond to the $A$ and $B$ phases, respectively.

If $\ket{u_n (k_x, t)}$ is an eigenstate with an energy eigenvalue $E_{n} (k_x, t)$,
then the  wave function
$\hat{G} \ket{u_n (k_x, t)}$ is also an eigenstate of the Hamiltonian at time $ t + \frac{T}{2} $ with the same energy eigenvalue $E_{n} (k_x, t)$
due to Eq.~(\ref{seq:adiabatic_SSH_01}):
\begin{eqnarray} 
\mathsf{H}_{\text{SSH}} \left[ \frac{T}{2} +t \right]  \hat{G} \ket{u_n (k_x, t)}
&=&  E_{n} (t) \hat{G} \ket{u_n (k_x, t)},
\\
\Ket{u_n \left(k_x,\frac{T}{2} + t\right)} & \propto & \hat{G} \ket{u_n (k_x, t)}.
\label{seq:SSH_soliton_charge_class_I}
\end{eqnarray}
Similarly, if $\ket{u_n (k_x, \frac{T}{4} - t )}$ is an eigenstate with an energy eigenvalue $E_{n} (k_x, \frac{T}{4}-t)$,
then the wave function
$\hat{ \Gamma}_{\text{II}} \ket{u_n (k_x, \frac{T}{4} - t )}$ is also an eigenstate of the Hamiltonian at time $ \frac{T}{4} + t $ 
with the opposite energy eigenvalue $-E_{n} (k_x, \frac{T}{4} - t )$ due to Eq.~(\ref{seq:adiabatic_SSH_02}):
\begin{eqnarray} 
&&\mathsf{H}_{\text{SSH}} \left[ k_x, \frac{T}{4} +t \right]  \hat{ \Gamma}_{\text{II}} \Ket{u_n \left(k_x, \frac{T}{4} - t \right)}
\\
&& ~~~~~~~~~~~~= -E_{n} \left(k_x, \frac{T}{4} - t \right) \hat{ \Gamma}_{\text{II}} \Ket{u_n \left(k_x, \frac{T}{4} - t \right)}.\nonumber
\\
&& \Ket{u_{3-n} \left(k_x,\frac{T}{4} +t \right)} \propto \hat{ \Gamma}_{\text{II}} \Ket{u_n \left(k_x,\frac{T}{4} - t \right)} .
\label{seq:SSH_soliton_charge_class_II}
\end{eqnarray}

Now, we prove the relation between soliton charges using the class~I operator $\hat{G}$.
Let us consider the partial adiabatic process from $t=0$ to $t=\frac{T}{2}$ that corresponds to the  soliton $S$ interpolating from $A$ to $B$ phases.
Then the corresponding partial Chern number $C_{S}$ is given by
\begin{eqnarray}
C_{S} = \frac{i}{ 2 \pi} \sum_{n = \text{occ}}{\vphantom{\sum}} \int_{\text{BZ}}
\int_{0}^{\frac{T}{2}}  dt ~ \Omega_n (\hat{I}, t),
\end{eqnarray}
where $\hat{I}$ is an identity operator.
Similarly, the partial adiabatic process from $t=\frac{T}{2}$ to $t=T$ corresponds to an antisoliton $ \bar S $ interpolating from $B$ to $A$ phases.
Then  one can show that the corresponding partial Chern number $C_{\bar S}$ is equal to $C_{S}$ using the glide operator $\hat{G}$:
\begin{eqnarray*}
C_{\bar S }
& = &  \frac{i}{ 2 \pi} \sum_{n = \text{occ}}{\vphantom{\sum}} \int_{\text{BZ}}  \int_{0}^{\frac{T}{2}}  dt
~ \Omega_n \left(\hat{I}, t+\frac{T}{2}\right),
\\
& = &  \frac{i}{ 2 \pi} \sum_{n = \text{occ}}{\vphantom{\sum}} \int_{\text{BZ}} \int_{0}^{\frac{T}{2}}  dt
~ \Omega_n (\hat{G}, t),
\\
& = & C_{S},
\end{eqnarray*}
where Eq.~(\ref{seq:SSH_soliton_charge_class_I}) is used from the first to the second lines
and  $\hat{G}^{\dagger} \hat{G} = 1$  and  $\Braket{  u_n(k_x,t ) |  u_n(k_x, t )} = 1$ are used from the second
to the last lines.
Thus, $Q_S = Q_{\bar{S}}$.

Next, we prove the relation between soliton charges using the class~II operator $\hat{\Gamma}_{\text{II}}$.
In this case, let us consider the partial adiabatic process from $t=\frac{T}{2}$ to $t=0$ that corresponds to an antisoliton $ \bar S $ interpolating from $B$ to $A$ ground states.
Then the corresponding partial Chern number $C_{\bar S}$ is equal to $- C_{S}$:
\begin{eqnarray*}
C_{\bar S} 
&= &\frac{i}{ 2 \pi} \sum_{n = \text{occ}}{\vphantom{\sum}} \int_{\text{BZ}}  \int_{+\frac{T}{4}}^{-\frac{T}{4}}  dt
~\Omega_n \left(\hat{I}, t+\frac{T}{4}\right),
\\
&= &\frac{i}{ 2 \pi}\sum_{n = \text{unocc}}{\vphantom{\sum}}
\int_{\text{BZ}}  \int_{+\frac{T}{4}}^{-\frac{T}{4}}  dt
~\Omega_n \left(\hat{\Gamma}_{\text{II}}, -t+\frac{T}{4}\right),
\\
&= &\frac{i}{ 2 \pi}\sum_{n = \text{unocc}}{\vphantom{\sum}}
\int_{\text{BZ}}  \int_{-\frac{T}{4}}^{+\frac{T}{4}}  dt
~\Omega_n \left(\hat{I}, t+\frac{T}{4}\right),
\\
&= & - C_{S},
\end{eqnarray*}
where Eq.~(\ref{seq:SSH_soliton_charge_class_II}) is used from the first to the second lines
and $ (\hat{\Gamma}^{(1)}_{\text{II}} )^{\dagger} \hat{\Gamma}^{(1)}_{\text{II}}  = 1$, $\Braket{  u_n(k_x,t ) |  u_n(k_x, t )} = 1$,  and $t \rightarrow - t$ are used from the second to the third lines.
From the third to the last lines, we use the fact that the total sum over the occupied and unoccupied states is zero.
Thus, $Q_S = - Q_{\bar{S}}$.
In a similar way, one can prove $Q_{S} = -Q_{S}$, $Q_{\bar{S}} = -Q_{\bar{S}}$ using class III operator.

Similarly, for the RM model, 
the soliton charge $Q_{S^*}$ and the antisoliton charge $Q_{\bar S^*}$
can be obtained from the adiabatic processes $A^* \rightarrow B^*$ and $B^* \rightarrow A^*$, respectively [see Figs.~\ref{fig:solitons}(f) and \ref{fig:adiabatic_evolution}(b1)].
The results are summarized in Table \ref{table:topological_charge}.

\subsection{Double-chain model}

Like the single-chain model, we show the role of the glide reflection operator $\hat{G}_y$ (class~I)
and the  nonsymmorphic chiral operator $\hat{\Gamma}_{\text{II}}^{(i)}$ (class~II) 
on the topological charges of chiral solitons in the DC model.
For the DC model, there are three types of the cyclic adiabatic evolutions depending on the type of chiral solitons 
as shown in Figs.~\ref{fig:solitons}(g)--\ref{fig:solitons}(i) and \ref{fig:adiabatic_evolution}(c1)--\ref{fig:adiabatic_evolution}(e1).
Without loss of generality, 
we choose the adiabatic processes along the straight lines in the order parameter space.
These adiabatic processes respect the transformation properties which are imposed by $\hat{G}_y$ and $\hat{\Gamma}_{\text{II}}^{(i)}$ .
For example, the phase-space Hamiltonians for the successive four RC and LC solitons
satisfy the following relations:
\begin{eqnarray}
\mathsf{H}_{\text{DC}}\left[ k_x, \frac{T}{4} +t \right]
& = &   \hat{G}_y  \mathsf{H}_{\text{DC}} [ k_x, t ] \hat{G}_y^{-1} \label{eq:Adiabatic_Glide_Reflection}
\end{eqnarray}
and
\begin{eqnarray}
&& \mathsf{H}_{\text{DC}}\left[ k_x, \frac{4m-3}{8}T +t\right]  \label{eq:Adiabatic_DC_NSC}
\\
&& ~~~~~~~~~~
= - \hat{\Gamma}^{(1)}_{\text{II}} \mathsf{H}_{\text{DC}} \left[ k_x, \frac{4m-3}{8}T-t \right] (\hat{\Gamma}^{(1)}_{\text{II}}  )^{-1}, \nonumber
\\
&& \mathsf{H}_{\text{DC}}\left[ k_x, \frac{4m-1}{8}T +t\right] 
\\
&& ~~~~~~~~~~
=- \hat{\Gamma}^{(2)}_{\text{II}}    \mathsf{H}_{\text{DC}} \left[ k_x, \frac{4m-1}{8}T-t \right] (\hat{\Gamma}^{(2)}_{\text{II}}  )^{-1}, \nonumber
\end{eqnarray}
where $m \in \mathbb{Z},~~ t \in [0, \frac{1}{8}T]$.

If $\ket{u_n (k_x, t)}$ is an eigenstate with an energy eigenvalue $E_{n} (k_x, t)$,
then the wave function
$ \hat{G}_y  \ket{u_n (k_x, t)}$ is also an eigenstate of the Hamiltonian at time $ t + \frac{T}{4} $  with the same energy eigenvalue $E_{n} (k_x, t)$
due to Eq.~(\ref{eq:Adiabatic_Glide_Reflection}):
\begin{eqnarray}  \label{seq:double_soliton_charge_class_I}
&& \mathsf{H}_{\text{DC}} \left[ k_x, t +\frac{T}{4}\right]  \hat{G}_y \Ket{u_n \left(k_x, t+\frac{T}{4}\right)}
\\
&& =E_{n}(k_x, t) \hat{G}_y \Ket{u_n \left(k_x, t+\frac{T}{4}\right)}.
\nonumber
\end{eqnarray}
Similarly, if $\ket{u_n (k_x, \frac{T}{8} - t )}$ is an eigenstate with an energy eigenvalue $E_{n} (k_x, \frac{T}{8}-t)$,
then the wave function
$\hat{ \Gamma}_{\text{II}}^{(1)} \ket{u_n (k_x, \frac{T}{8} - t )}$ is also an eigenstate of the Hamiltonian at time $ \frac{T}{8} + t $  
with the opposite energy eigenvalue $-E_{n} (k_x, \frac{T}{8} - t )$
due to Eq.~(\ref{eq:Adiabatic_DC_NSC}):
\begin{eqnarray} \label{seq:double_soliton_charge_class_II}
&& \mathsf{H}_{\text{DC}} \left[ k_x, \frac{T}{8} + t \right] \hat{ \Gamma}_{\text{II}}^{(1)} \Ket{u_n \left(k_x, \frac{T}{8} - t \right)}
\\
&& =- E_{n} \left(k_x, \frac{T}{8} - t \right) \hat{ \Gamma}_{\text{II}}^{(1)} \Ket{u_n \left(k_x, \frac{T}{8} - t \right)}. \nonumber
\end{eqnarray}

Now, we prove the relation among soliton charges using the glide reflection operator $\hat{G}_y$.
Let us consider the partial adiabatic process from $t=0$ to $t=\frac{T}{4}$ that corresponds to the RC soliton $S^R_1$ 
interpolating from $AA$ to $BA$ phases.
Then the corresponding partial Chern number $C_{S^R_1}$ is defined as
\begin{eqnarray}
C_{S^R_1} = \frac{i}{ 2 \pi} \sum_{n = \text{occ}}{\vphantom{\sum}} \int_{\text{BZ}} \int_{0}^{\frac{T}{4}}  dt
~ \Omega_n (\hat{I}, t),
\end{eqnarray}
where $\hat{I}$ is an identity operator.

Next let us consider the partial adiabatic process from $t=\frac{T}{4}$ to $t=\frac{T}{2}$ that corresponds to the RC soliton $S^R_2$ 
interpolating from $BA$ to $BB$ phases.
Then the corresponding partial Chern number $C_{S^R_2}$ is equal to $C_{S^R_1}$:
\begin{eqnarray*}
C_{S^R_2}
& = &  \frac{i}{ 2 \pi} \sum_{n = \text{occ}}{\vphantom{\sum}} \int_{\text{BZ}} \int_{0}^{\frac{T}{4}}  dt
~\Omega_n \left(\hat{I}, t+\frac{T}{4}\right),
\\
& = &  \frac{i}{ 2 \pi} \sum_{n = \text{occ}}{\vphantom{\sum}} \int_{\text{BZ}}  \int_{0}^{\frac{T}{4}}  dt
~\Omega_n (\hat{G}, t),
\\
& =& C_{S^R_1},
\end{eqnarray*}
where Eq.~(\ref{seq:double_soliton_charge_class_I}) is used from the first to the second lines
and  $\hat{G}_y^{\dagger} \hat{G}_y=1$  and  $\Braket{  u_n(k_x,t ) |  u_n(k_x, t )} = 1$ are used from the second to the last lines.
Thus, $Q_{S^R_1} = Q_{S^R_2}$.
Similarly, one can prove that topological charges for the chiral solitons of the same chirality are the same:
\begin{eqnarray}
&& Q_{S^k_1} =  Q_{S^k_2} =  Q_{S^k_3}  =  Q_{S^k_4}, \label{seq:chrial_charge_equiv_01} 
\end{eqnarray}
where $k=R, L, A$.

Next, we prove the relation among soliton charges using the nonsymmorphic chiral operator $\hat{\Gamma}^{(i)}_{\text{II}}$.
Let us consider the partial adiabatic process from $t=0$ to $t=\frac{T}{4}$ that corresponds to the RC soliton $S_1^R$ interpolating from $AA$ to $BA$ phases.
Then the corresponding partial Chern number $C_{S_1^R}$ can be written as
\begin{eqnarray*}
C_{S_1^R} 
&= &\frac{i}{ 2 \pi} \sum_{n = \text{occ}}{\vphantom{\sum}} \int_{\text{BZ}} \int_{-\frac{T}{8}}^{+\frac{T}{8}} dt
~\Omega_n \left(\hat{I}, t+\frac{T}{8}\right).
\end{eqnarray*}
Let us consider the partial adiabatic process from $t=\frac{T}{4}$ to $t=0$ that corresponds to the LC soliton $S_4^L$ interpolating from $BA$ to $AA$ phases.
Then the corresponding partial Chern number $C_{S_4^L}$ is equal to $-C_{S_1^R}$:
\begin{eqnarray*}
C_{S_4^L} 
&= &\frac{i}{ 2 \pi} \sum_{n = \text{occ}}{\vphantom{\sum}} \int_{\text{BZ}}   \int_{+\frac{T}{8}}^{-\frac{T}{8}}  dt
~\Omega_n \left(\hat{I}, t+\frac{T}{8}\right),
\\
&= &\frac{i}{ 2 \pi} \sum_{n = \text{unocc}}{\vphantom{\sum}} \int_{\text{BZ}}   \int_{+\frac{T}{8}}^{-\frac{T}{8}}  dt
~\Omega_n \left(\hat{\Gamma}^{(1)}_{\text{II}}, -t+\frac{T}{8}\right),
\\
&= &\frac{i}{ 2 \pi} \sum_{n = \text{unocc}}{\vphantom{\sum}} \int_{\text{BZ}}   \int_{-\frac{T}{8}}^{+\frac{T}{8}}  dt
~\Omega_n \left(\hat{I}, t+\frac{T}{8}\right),
\\
&= & - C_{S_1^R},
\end{eqnarray*}
where Eq.~(\ref{seq:double_soliton_charge_class_II}) is used from the first to the second lines
and 
$ (\hat{\Gamma}^{(1)}_{\text{II}} )^{\dagger} \hat{\Gamma}^{(1)}_{\text{II}}  = 1$, $\Braket{  u_n(k_x,t ) |  u_n(k_x, t )} = 1$,  and $t \rightarrow - t$ are used from the second to the third lines.
From the third to the last lines, we use the fact that the total sum over the occupied and unoccupied states is zero.
Thus, $ Q_{S_1^R} = - Q_{S_2^L} $.
Similarly, one can show that the other pairs of RC and LC solitons have the opposite topological charges:
\begin{eqnarray}
Q_{S_i^R} &=& - Q_{S_{5-i}^L} ~~ ( i= 1, 2, 3, 4) ~~ \text{for} ~~ \hat{\Gamma}^{(1)}_{\text{II}}, \label{seq:chrial_charge_PH_01} \\
Q_{S_i^R} &=& - Q_{S_{i+(-1)^{i+1}}^L} ~~ ( i=  1, 2, 3, 4 ) ~~ \text{for} ~~ \hat{\Gamma}^{(2)}_{\text{II}}.  \label{seq:chrial_charge_PH_02}
\end{eqnarray}
Also, one can show that the other pairs of two different AC solitons have the opposite topological charges:
\begin{eqnarray}
Q_{S_i^A} &=& - Q_{S_{i+1}^A} ~~ (i = 1, 3 ) ~~ \text{for} ~~ \hat{\Gamma}^{(1)}_{\text{II}}, \label{seq:chrial_charge_PH_03}\\
Q_{S_i^A} &=& - Q_{S_{5-i}^A} ~~ (i = 2, 4 ) ~~ \text{for} ~~ \hat{\Gamma}^{(2)}_{\text{II}}.  \label{seq:chrial_charge_PH_04}
\end{eqnarray}
In a similar way, one can prove the other relations using class III operators.

\section{Wannier charge center} \label{appendix:wannier charge center}
In this appendix, we discuss the relations between the Wannier charge centers~\cite{marzari1997} and the Berry phases in the quasi-1D systems.
Based on the relations, the Wannier charge centers are numerically calculated for the SSH, RM and DC models [see Figs.~\ref{fig:adiabatic_evolution}(a1)--\ref{fig:adiabatic_evolution}(e1)]. 
The Wannier state localized at the $j$th unit cell is given by 
\begin{eqnarray}
\ket{ W_n (j) } = \frac{1}{\sqrt{N}} \sum_{k_x} e^{- i k_x j} \ket {\psi_n (k_x)},
\end{eqnarray}
where $n$ is the band index, $\ket {\psi_n(k_x)}$ is a Bloch state, and the summation is done over
$ k_x = m \frac{2 \pi}{N a_0}$ with $ m = 1, \ldots, N $.
For the single-chain model, the position operator $\hat x$ is given by
\begin{eqnarray}
\hat x = \sum_{m=1}^{N} \sum_{n=1}^{2} m \left (   \ket{m, a_n} \bra{m, a_n} \right ) ,
\end{eqnarray}
where $\ket{m, a_n}$ is the state localized at atom $a_n$ in the $m$th unit cell. Here $ a_n = (a,b)$.
Then the Wannier center of the $j$th cell is given by
\begin{eqnarray}
\Braket{W_n (j) |\hat x | W_n (j)}&=&  j + \frac{i}{2\pi} \int_{0}^{ \frac{2 \pi}{a_0} } d k_x \Braket{ u_n (k_x) | \partial_{k_x} u_n (k_x) }, \nonumber
 \\
& = & j + \frac{ \gamma}{2\pi}, 
\end{eqnarray}
where $\gamma$ is the Berry phase.

However, for the DC model, the normalized position operator $\hat x$ is given by
\begin{eqnarray}
\hat x = \frac{1}{2} \sum_{m=1}^{N}  \sum_{n=1}^{4} m \left (   \ket{m, a_n} \bra{m, a_n} \right ) ,
\end{eqnarray}
where $\ket{m, a_n}$ is the state localized at atom $a_n$ in the $m$th unit cell. Here $ a_n= (a,b,c,d)$.
In this case, the Wannier center of the $j$th cell is given by
\begin{eqnarray}
\Braket{W_n (j) |\hat x | W_n (j)}&=&  j + \frac{i}{4\pi} \int_{0}^{ \frac{2 \pi}{a_0} } d k_x \Braket{ u_n (k_x) | \partial_{k_x} u_n (k_x) }, \nonumber
\\
&=&  j + \frac{1}{2}\frac{ \gamma}{2\pi}. 
\end{eqnarray}

\section{2D Effective Hamiltonians} \label{appendix:2D effective Hamiltonian}

\subsection{Tight-binding Hamiltonian}
To understand the topological properties and chiral nature of the chiral solitons,
we take into account an cyclic adiabatic evolution of a 1D Hamiltonian $H_{\text{1D}}(k_x, t)$
and we extend the 1D system into a 2D system
by substituting the time-evolution as momentum $k_y$ in an extra dimension.
Then we construct the 2D Hamiltonian $H_{\text{2D}}(k_x, k_y)$
such that $H_{\text{2D}}(k_x, k_y=0) = H_{\text{1D}}(k_x, t=0)$ and $H_{\text{2D}}(k_x, k_y=2 \pi) = H_{\text{1D}}(k_x, t=T)$.

For the single-chain model, the 2D tight-binding Hamiltonian $H_{\text{2D}}^{\text{single}}$ is given by
\begin{eqnarray*}
&& H_{\text{2D}}^{\text{single}}  = t_0 \sum _{n_x, n_y} c_{n_x+1 , n_y}^{\dagger} c_{n_x, n_y} + \text{H.c.} \\
&&~~~~~~~~~~~~~~~
+  m_z     \sum _{n_x, n_y}  (-1)^{n_x+1} c_{n_x , n_y}^{\dagger} c_{n_x, n_y}
	+ H_{\text{adiabatic}}, \\
&& H_{\text{adiabatic}} 
	= \frac{\Delta_0}{4} \sum _{n_x, n_y}  
	(-1)^{n_x+1} \left  ( c_{n_x , n_y}^{\dagger} c_{n_x+1, n_y+1} \right. 
	\\
&&~~~~~~~~~~~~~~~
\left. + c_{n_x , n_y}^{\dagger} c_{n_x-1, n_y-1}\right )  + \text{H.c.} ,
\end{eqnarray*}
where $n_x$ and $n_y$ indicate the lattice sites along the original chain direction and the cyclic direction for the phase evolution, respectively. 
When $m_z = 0$ ($m_z \neq 0$), the tight-binding Hamiltonian corresponds to the adiabatic evolution for the SSH (RM) model.

For the double-chain model, the 2D tight-binding Hamiltonian $H_{\text{2D}}^{\text{DC}}$ is given by
\begin{eqnarray*}
&&H_{\text{2D}}^{\text{DC}} = H^{(1)}_{\text{2D}}  + H^{(2)}_{\text{2D}}  + H_{\text{coupling}},  \\
&&H^{(i)}_{\text{2D}} 
=  t_0 \sum _{n_x, n_y} \left[ c_{n_x+1 , n_y}^{(i) \dagger} c_{n_x, n_y}^{(i)} + \text{H.c.} \right] + H_{\text{adiabatic}} ^{(i)} , \\
&& H_{\text{adiabatic}} ^{(i)}
= \frac{ \tilde \Delta_i}{4} \sum _{n_x, n_y}  
               (-1)^{n_x+1} \left  [ c_{n_x , n_y}^{(i)\dagger} c_{n_x+1, n_y+1}^{(i)} \right. 
               \\
&&~~~~~~~~~~~~~~~~~~~
    \left.  + c_{n_x , n_y}^{(i)\dagger} c_{n_x-1, n_y-1}^{(i)} \right ] + \text{H.c.} , \\
&& H_{\text{coupling}} = \delta  t_0 \sum_{n_x, n_y} 
               \left[ c_{n_x , n_y}^{(1) \dagger} c_{n_x, n_y}^{(2)} +  c_{n_x , n_y}^{(1) \dagger} c_{n_x+1, n_y}^{(2)}\right] + \text{H.c.}
\end{eqnarray*}
For the cyclic evolution of $AA \rightarrow BA \rightarrow BB \rightarrow AB \rightarrow AA$, $( \tilde \Delta_1, \tilde \Delta_2 )$ is set to be $\Delta_0 (1+i, 1-i)$.
For the reversed path, $( \tilde \Delta_1, \tilde \Delta_2 )=\Delta_0 (1-i, 1+i)$.
For the evolution of $AA \rightarrow BB \rightarrow AA$, $( \tilde \Delta_1, \tilde \Delta_2 )= \Delta_0 (1, 1)$.

\subsection{Bloch Hamiltonian and time-reversal symmetry for the single-chain model}
From the tight-binding Hamiltonian, we construct the 2D Bloch Hamiltonian $\mathcal{H}_{\text{single}}^{\text{2D}}  (k_x, k_y) $   
for the SSH and RM models, which is given by
\begin{eqnarray*}
	\mathcal{H}_{\text{single}}^{\text{2D}} (k_x, k_y) 
	&=& 2t_0 \cos \left( \frac{k_x a_0}{2} \right) \sigma_x 
	\\
	&&-\Delta_0 \cos ( k_y b )   \sin \left( \frac{ k_x a_0}{2}\right) \sigma_y  +m_z \sigma_z.
\end{eqnarray*}
This Hamiltonian satisfies the time-reversal symmetry regardless of the sublattice symmetry breaking:
\begin{eqnarray}
\hat{\mathcal{T}} \mathcal{H}_{\text{single}}^{\text{2D}} (k_x, k_y) \hat{\mathcal{T}} ^{ -1 } 
= \mathcal{H}_{\text{single}}^{\text{2D}} (k_x, k_y),
\end{eqnarray} 
where $\hat{\mathcal{T}}  = \hat{ K } \otimes ( \mathbf{k} \rightarrow -\mathbf{k} )$ is the 2D time-reversal operator.
Therefore, the total Chern number is zero.

\subsection{Bloch Hamiltonian and time-reversal symmetry for the double-chain model}
From the tight-binding Hamiltonian, we construct the 2D Bloch Hamiltonian
$\mathcal{H}_{\text{DC}}^{\text{2D}}   ( k_x , k_y )$   
for the double-chain model, which is given by
\begin{eqnarray}
&& \mathcal{H}_{\text{DC}}^{\text{2D}}   ( k_x , k_y ) =
\begin{pmatrix}
\mathcal{H}_{1} & \mathcal{H}_{12}\\
\mathcal{H}_{21} & \mathcal{H}_{2}
\end{pmatrix},
\end{eqnarray}
with
\begin{eqnarray*}
&& \mathcal{H}_{i}   = ( 2t_0 \cos ( k_x a_0/2 ) , {\color{blue}-}\Delta^{(i)} \sin ( k_x a_0 / 2 ), 0 ) \cdot \boldsymbol{\sigma},
\\
&& \mathcal{H}_{12} = \mathcal{H}^{\dagger}_{21} = \delta t_0 (e^{- i k_x a_0/4} 1_{2 \times 2} + e^{i k_x a_0/4} \sigma_x).
\end{eqnarray*}
For the RC and LC solitons, 
\begin{eqnarray}
	\Delta^{(1)} & = & \Delta_0 ( \cos  k_y  - \sin  k_y  ), \\
	\Delta^{(2)} & = & \pm \Delta_0 ( \cos  k_y  + \sin  k_y  ),
	\\ \nonumber
\end{eqnarray}
where $+$ and $-$ correspond to the RC and LC solitons, respectively.
Therefore, the time-reversal symmetry is broken:
\begin{eqnarray}
\hat{\mathcal{T}} \mathcal{H}_{\text{DC}}^{\text{2D}} (k_x, k_y) \hat{\mathcal{T}} ^{ -1 } 
\neq  \mathcal{H}_{\text{DC}}^{\text{2D}} (k_x, k_y).
\end{eqnarray} 
Therefore, the 2D effective Hamiltonians for the RC and LC solitons have non-zero total Chern numbers.

On the other hand, the 2D Bloch Hamiltonians for the AC solitons have the time-reversal symmetry.
For example, for the evolution of $AA\rightarrow BB\rightarrow AA$,
\begin{eqnarray}
	( \Delta^{(1)} , \Delta^{(2)} ) =  \Delta_0 ( \cos  k_y  ,  \cos   k_y  ).
\end{eqnarray}
Then the Hamiltonian  has the time-reversal symmetry:
\begin{eqnarray}
\hat{\mathcal{T}} \mathcal{H}_{\text{DC}}^{\text{2D}} (k_x, k_y) \hat{\mathcal{T}} ^{ -1 } 
=  \mathcal{H}_{\text{DC}}^{\text{2D}} (k_x, k_y).
\end{eqnarray} 
Therefore, the 2D Bloch Hamiltonians for the AC solitons have a zero total Chern number.

\begin{acknowledgments} 
We thank S.-H. Lee, K.-S. Kim, and H. W. Yeom for discussions in the early stages of this work and Myungjun Kang for proofreading.
S.-H.H., S.-W.K., and S.C. were supported by the National Research Foundation (NRF) of Korea through Basic Science Research Programs (NRF-2018R1C1B6007607), the research fund of Hanyang University (HY-2017), and the POSCO Science Fellowship of POSCO TJ Park Foundation.
S.-G.J. and T.-H.K. were supported by the Institute for Basic Science (IBS-R014-D1) and the NRF grant funded by the Korea government (MSIT) (NRF-2016K1A4A4A01922028, NRF-2018R1A5A6075964).
\end{acknowledgments}

\bibliographystyle{apsrev4-1}
%

\end{document}